\documentclass[aps,prl,reprint,superscriptaddress,floatfix,longbibliography]{revtex4-2}

\usepackage{amsmath}
\usepackage{amsfonts}
\usepackage{newtxtext}
\usepackage{newtxmath}
\usepackage{mathtools}
\usepackage[mathcal]{eucal}
\usepackage{bbm}
\usepackage{bm}
\usepackage{braket}
\usepackage{ulem}

\usepackage{microtype}

\usepackage{tabularray}
\usepackage[table,dvipsnames]{xcolor}
\usepackage[percent]{overpic}
\UseTblrLibrary{booktabs}
\usepackage{array}
\usepackage[caption=false]{subfig}

\usepackage[
    bookmarksnumbered,
    pdfpagelabels=true,
    plainpages=false
    ]{hyperref}
\usepackage[capitalize]{cleveref}
\definecolor{cblue}{RGB}{55,126,184}
\hypersetup{
    colorlinks=true,
    linkcolor=cblue,
    citecolor=cblue,
    urlcolor=cblue
    }

\usepackage{tikz}
\tikzset{
    mpo/.style={
        draw,
        shape          = rectangle,
        fill           = white,
        inner sep      = 0.6mm,
        minimum width  = 0.6cm,
        minimum height = 0.6cm
    }
}

\usetikzlibrary{decorations.pathmorphing} 

\usepackage{pdfpages}

\makeatletter
\AtBeginDocument{\let\LS@rot\@undefined}
\makeatother

\usepackage{pgffor}

\DeclareMathOperator{\Rep}{Rep}
\DeclareMathOperator{\Tr}{Tr}
\renewcommand{\Re}{\mathrm{Re}}
\renewcommand{\Im}{\mathrm{Im}}

\newcommand{\Xl }[1]{
    \raisebox{2.3mm}{\scalebox{.7}{$\rightarrow$}} 
    \mkern-15mu 
    X^{\hspace{.3mm}#1}
    }
\newcommand{\Xr }[1]{
    \raisebox{2.3mm}{\scalebox{.7}{$\leftarrow$}} 
    \mkern-14mu 
    X^{\hspace{.3mm}#1}
    }

\newcommand{\xl }[1]{
    \raisebox{2.5mm}{\scalebox{.7}{$\rightarrow$}} 
    \mkern-13mu 
    \mathcal{X}^{\hspace{.3mm}\raisebox{.7mm}{$\scriptstyle #1$}}
    }
\newcommand{\xr }[1]{
    \raisebox{2.5mm}{\scalebox{.7}{$\leftarrow$}} 
    \mkern-12mu 
    \mathcal{X}^{\hspace{.3mm}\raisebox{.7mm}{$\scriptstyle #1$}}
    }

\newcommand{\z}{\mathcal{Z}}

\newcommand{\Ol}[1]{
    \raisebox{2.3mm}{\scalebox{.7}{$\rightarrow$}} 
    \mkern-15mu 
    \mathcal{O}^{\hspace{.3mm}#1}
    }
\newcommand{\Or}[1]{
    \raisebox{2.3mm}{\scalebox{.7}{$\leftarrow$}} 
    \mkern-14mu 
    \mathcal{O}^{\hspace{.3mm}#1}
    }

\makeatletter 
\renewcommand\onecolumngrid{
\do@columngrid{one}{\@ne}%
\def\set@footnotewidth{\onecolumngrid}
\def\footnoterule{\kern-6pt\hrule width 1.5in\kern6pt}%
}
\makeatother

\makeatletter
\newcommand{\raisemath}[1]{\mathpalette{\raisem@th{#1}}}
\newcommand{\raisem@th}[3]{\raisebox{#1}{$#2#3$}}
\makeatother

\renewcommand{\emph}{\textit}

\begin{document}

\title{Spontaneously Broken Non-Invertible Symmetries in Transverse-Field Ising Qudit Chains}

\author{Kristian Tyn Kai Chung}
\email[Corresponding author: ]{ktchung.phys@gmail.com}
\affiliation{Max Planck Institute for the Physics of Complex Systems, N\"othnitzer Strasse 38, 01187 Dresden, Germany}
\affiliation{Department of Physics and Astronomy, Rice University, Houston, Texas 77005, USA}

\author{Umberto Borla}
\affiliation{Max Planck Institute of Quantum Optics, 85748 Garching, Germany}
\affiliation{Munich Center for Quantum Science and Technology (MCQST), 80799 Munich, Germany}
\affiliation{Racah Institute of Physics, The Hebrew University of Jerusalem, Givat Ram, Jerusalem 91904, Israel}

\author{Andriy H. Nevidomskyy}
\affiliation{Department of Physics and Astronomy, Rice University, Houston, Texas 77005, USA}
\affiliation{Rice Center for Quantum Materials and Advanced Materials Institute, Rice University, Houston, Texas 77005, USA}
\affiliation{Division of Condensed Matter Physics and Materials Science, Brookhaven National Laboratory, Upton, NY 11973-5000, USA}

\author{Sergej Moroz}
\affiliation{Department of Engineering and Physics, Karlstad University, Karlstad, Sweden}
\affiliation{Nordita, KTH Royal Institute of Technology and Stockholm University, Stockholm, Sweden}

\date{\today}

\begin{abstract}
    Recent developments have revealed that symmetries need not form a group, but instead can be non-invertible. Here we use analytical arguments and numerical evidence to illuminate how spontaneous symmetry breaking of a non-invertible symmetry is similar yet distinct from ordinary, invertible, symmetry breaking. We consider one-dimensional chains of group-valued qudits, whose local Hilbert space is spanned by elements of a finite group $G$ (reducing to ordinary qubits when $G=\mathbbm{Z}_2$). We construct Ising-type transverse-field Hamiltonians with Rep($G$) symmetry whose generators multiply according to the tensor product of irreducible representations (irreps) of the group $G$. For non-Abelian $G$, the symmetry is non-invertible. In the symmetry broken phase there is one ground state per irrep on a closed chain. The symmetry breaking can be detected by local order parameters but, unlike the invertible case, different ground states have distinct entanglement patterns. We show that for each irrep of dimension greater than one the corresponding ground state exhibits string order, entanglement spectrum degeneracies, and has gapless edge modes on an open chain -- features usually associated with symmetry-protected topological order. Consequently, domain wall excitations behave as one-dimensional non-Abelian anyons with non-trivial internal Hilbert spaces and fusion rules. Our work identifies properties of non-invertible symmetry breaking that existing quantum hardware can probe.   
\end{abstract}

\maketitle

\emph{Introduction}---Symmetries provide powerful organizing principles that dictate the form of physical laws, constrain dynamics, and enable the classification of phases of matter.
Due to Landau~\cite{landauTheoryPhaseTransitions1936}, phases of matter may be classified according to patterns of spontaneous symmetry breaking (SSB), as characterized by local order parameters which are organized into charged multiplets. The prototypical example is the spontaneous breaking of $\mathbbm{Z}_2$ symmetry in the transverse field Ising model (TFIM) \cite{liebTwoSolubleModels1961} or its classical 2D equivalent~\cite{peierlsIsingsModelFerromagnetism1936,onsagerCrystalStatisticsTwoDimensional1944a,yangSpontaneousMagnetizationTwoDimensional1952,kasteleynDimerStatisticsPhase1963}. 
At the next level, unbroken symmetries can protect non-trivial entanglement in the ground state wavefunction of a many-body system, called a symmetry protected topological phase (SPT)~\cite{senthilSymmetryProtectedTopologicalPhases2015, zengQuantumInformationMeets2019}, which are characterized instead by a non-local string order parameter~\cite{dennijsPrerougheningTransitionsCrystal1989, kennedyHiddenSymmetryBreaking1992,pollmannSymmetryProtectionTopological2012,pollmannDetectionSymmetryprotectedTopological2012,elseHiddenSymmetrybreakingPicture2013,verresenOnedimensionalSymmetryProtected2017}. 
In recent years the notion of symmetry has expanded greatly under the broad umbrella of the ``generalized symmetries paradigm''~\cite{gaiottoGeneralizedGlobalSymmetries2015,mcgreevyGeneralizedSymmetriesCondensed2023}.
Whereas ordinary, invertible, symmetries are captured by their group structure, it is now understood that symmetries can instead be non-invertible (technically, described by a fusion category structure)~\cite{petkovaGeneralisedTwistedPartition2001,frohlichKramersWannierDualityConformal2004,frohlichDualityDefectsRational2007,aasen2016topological,aasenTopologicalDefectsLattice2020, shaoWhatsDoneCannot2023,schafer-namekiICTPLecturesNonInvertible2023,bhardwajCategoricalLandauParadigm2024}.
The past few years have seen an explosion of activity in the study of such non-invertible symmetries, with particular emphasis on their formal mathematical underpinnings and field theory manifestations~\cite{cordovaSnowmassWhitePaper2022,bhardwajUnifyingConstructionsNoninvertible2023,bhardwajClubSandwichGapless2025,roumpedakisHigherGaugingNoninvertible2023,bhardwajGeneralizedChargesPart2025,bhardwajGappedPhasesNoninvertible2025,  hsinFractionalizationCosetNonInvertible2024,bhardwajHasseDiagramsGapless2025,bartschNoninvertibleSymmetriesHigher2024}.
Recently, more attention is being paid to condensed-matter-relevant lattice systems~\cite{bhardwajLatticeModelsPhases2024,bhardwajIllustratingCategoricalLandau2025,10.21468/SciPostPhys.17.4.115,Lootens2025-ng,kobayashiProjectiveRepresentationsBogomolov2026,lootensDualitiesOneDimensionalQuantum2023,lootensDualitiesOneDimensionalQuantum2024,paceLatticeTdualityNoninvertible2025,choiNoninvertibleHigherformSymmetries2025,gorantlaTensorNetworksNoninvertible2025,seibergNoninvertibleSymmetriesLSMtype2024,seifnashriClusterStateNoninvertible2024,luRealizingTrialitypality2024,caoGlobalSymmetriesQuantum2025,10.21468/SciPostPhys.16.5.127, paceGaugingModulatedSymmetries2025,caoGeneratingLatticeNoninvertible2024,fechisinNoninvertibleSymmetryProtectedTopological2025,grafIsingDualreflectionInterface2024, 10.21468/SciPostPhys.18.1.028, 249m-m8wq,luStrangeCorrelatorString2025}.
However, there remains a large disconnect between these very technical developments and the broader condensed matter community, for whom invertible internal and spatial symmetries are textbook knowledge yet non-invertible symmetries remain largely mysterious.

\begin{figure}[b]
    \centering
    \includegraphics[width=\linewidth]{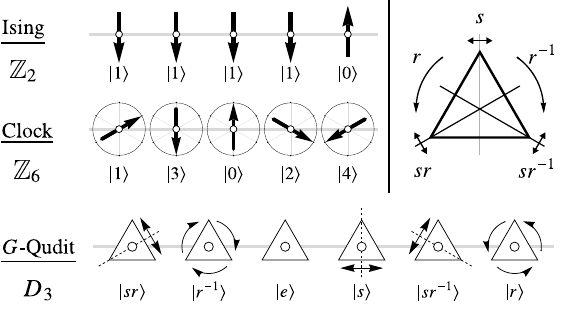}
    \caption{We consider Ising-like chains of $G$-qudits, whose internal states are labeled by the elements of a finite group~$G$. They reduce to the transverse-field Ising model when $G=\mathbbm{Z}_2$ and clock models when $G=\mathbbm{Z}_N$, but can harbor non-invertible Rep($G$) symmetries when $G$ is non-Abelian. The smallest discrete non-Abelian group is the dihedral group $D_3$, generated by a rotation~$r$ and a reflection~$s$.}
    \label{fig:Ising-G-Qudits}
\end{figure}

The aim of this paper is to bridge that divide, by studying spontaneous breaking of a non-invertible symmetry in a model with a tensor product Hilbert space which is a direct generalization of the celebrated TFIM, see Fig. \ref{fig:Ising-G-Qudits}. 
We show how ground states with distinct entanglement structures arise and how seemingly disparate features---conventional local order and non‑local string order---coexist in harmony, protected by the non‑invertible symmetry. Our results offer guidance for ongoing analog and digital quantum simulations with qudits and open new avenues toward non‑Abelian quantum computation in one spatial dimension.

\emph{Ising Models: From Qubits to $G$-Qudits}---The TFIM is the workhorse of quantum many-body physics, and for our purposes is the canonical example of invertible symmetry breaking. The Hamiltonian is
\begin{equation}
    H_{\text{TFIM}} = - J \sum_{i } Z_i Z_{i+1}^{\dagger} - h \sum_i X_i + \text{h.c.},
    \label{eq:TFIM}
\end{equation}
where $X$ and $Z$ are Pauli operators, $J>0$ is the ferromagnetic Ising coupling and $h$ the transverse field strength. 
This Hamiltonian has an invertible $\mathbbm{Z}_2$ symmetry generated by the unitary operator $U = \prod_i X_i$.
At large field there is a unique symmetric gapped ground state, the paramagnetic product state of the symmetric superposition $\ket{0} + \ket{1}$ on every site. 
At small field there are two ferromagnetic gapped ground states, $\cramped{\ket{\bm{0}}\equiv \bigotimes_i\ket{0}_i}$ and $\cramped{\ket{\bm{1}}\equiv\bigotimes_i\ket{1}_i}$; since $U \ket{\bm 0} = \ket{\bm 1}$ and vice-versa, the $\mathbbm{Z}_2$ symmetry is spontaneously broken.
This symmetry breaking is detected by operators such as $Z_i$ which are charged under the symmetry, $U Z_i = -  Z_i U$, which serve as local order parameters. 
In the symmetry broken phase the gapped excitations are fermionic domain walls, which behave as deconfined quasiparticles in this 1D system~\cite{liebTwoSolubleModels1961}.

The natural generalization of the Ising $\mathbbm{Z}_2$ symmetry to an Abelian $\mathbbm{Z}_N$ symmetry is a clock model~\cite{fradkinDisorderVariablesParafermions1980}. 
We upgrade the two-state qubits with $N$-state qudits, with states $\ket{n}$, and generalize the Pauli operators to ``clock'' operator, $\cramped{Z^k \ket{n} = e^{i2\pi kn/N}\ket{n}}$, and ``shift'' operator $\cramped{X^m \ket{n} = \ket{n+m}}$, with addition modulo~$N$.
They satisfy
\begin{equation}
    Z^k X^m = e^{i2\pi k m/N} X^m Z^k.
    \label{eq:ZX_ZN}
\end{equation}
Using the same Hamiltonian, \cref{eq:TFIM}, the
global $\mathbbm{Z}_N$ symmetry is generated by $U = \prod_i X_i$. 
The trivial state at large field is the symmetric superposition $\sum_n \ket{n}$ on every site, and the symmetry broken states at small field are $\cramped{\ket{\bm{n}}\equiv \bigotimes_i \ket{n}_i}$. 
The symmetry generator exchanges ground states, $\cramped{U^m\ket{\bm{n}} = \ket{\bm{n}+\bm{m}}}$, and the $Z_j$ operators again serve as local order parameters, as they are charged under $U$ due to \cref{eq:ZX_ZN}, with $\cramped{\bra{\bm{n}}Z_j\ket{\bm{n}} = e^{2\pi in /N}}$.
The domain walls are parafermionic quasiparticles~\cite{fradkinDisorderVariablesParafermions1980,fendleyParafermionicEdgeZero2012}. 

We now consider how this generalizes to an arbitrary finite, and in particular non-Abelian, group $G$. 
Following the pattern, the local Hilbert space is spanned by states $\ket{g}$ for each group element $g\in G$, which we refer to as a $G$-qudit~\cite{SM}. 
The clock and shift operators generalize to~\cite{brellGeneralizedClusterStates2015,fechisinNoninvertibleSymmetryProtectedTopological2025,albertSpinChainsDefects2021,paceSPTLSMTheoremsProjective2025,kitaevFaulttolerantQuantumComputation2003}
\begin{equation}
    \Xl{h} \ket{g} = \ket{hg} , 
    \,\,\,
    \Xr{h}\ket{g} = \ket{gh^{-1}} , 
    \,\,\,
    Z^{\Gamma}_{\alpha\beta} \ket{g} = \Gamma_{\alpha\beta}^g \ket{g}.
\end{equation}
The non-Abelian shift operators are self-explanatory---they perform group multiplication, either from the left or right (which are inequivalent for non-Abelian groups).
The non-Abelian clock operators are labeled by an irreducible representation (irrep) $\Gamma$, where $\Gamma_{\alpha\beta}^{\smash{g}}$ is the \mbox{$d_{\Gamma}$-dimensional} unitary representation matrix of group element $g$.
The non-Abelian generalization of \cref{eq:ZX_ZN} is~\cite{SM}
\begin{equation}
    Z_{\alpha\beta}^{\Gamma} \,
    \Xl{g}  
    =  
    \Gamma_{
        \alpha\gamma
    }^{
        \smash{g}
    }
    \,
    \Xl{g}
    Z^{\Gamma}_{\gamma\beta}
    \,,
    \label{eq:ZX_G}
\end{equation}
with implied summation of the repeated index $\gamma$ (but not $g$).
These reduce to the Pauli operators for $G = \mathbbm{Z}_2$ and clock operators when $G = \mathbbm{Z}_N$. 

Using the $G$-qudits, the simplest $G$-symmetric transverse-field Hamiltonian analogous to \cref{eq:TFIM} is \cite{PhysRevB.98.245135,albertSpinChainsDefects2021}
\begin{equation}
    H_{G} = -J \sum_i \sum_{\Gamma} d_{\Gamma} \Tr[Z^{\overline \Gamma}_i \cdot Z^{\Gamma}_{i+1}] - h \sum_i \sum_g \Xr{g}_i + \text{h.c.},
    \label{eq:H_G}
\end{equation}
where in the first term the dot $(\cdot)$ indicates contraction of the neighboring indices of the $Z$ operators, while the trace indicates contraction of their outer indices, and the bar indicates $\smash{\cramped{\overline{\Gamma}^{\raisebox{-.8ex}{$\scriptstyle g$}} = \Gamma^{g^{-1}}}}$.
This Hamiltonian has a $G$ symmetry generated by invertible operators $\smash{U_g = \prod_i \Xl{g}_i}$.
It also has other symmetries irrelevant to the current discussion; its most general version is given in Appendix A.
The first term is a projector which is zero unless neighboring sites are in the same group element state, thus at small field the ground states are $\cramped{\ket{\bm{g}} \equiv \bigotimes_i \ket{g}_i}$, spontaneously breaking the $G$ symmetry. 
The second term serves to disorder the group elements, such that at large field the unique gapped ground state is the symmetric superposition $\sum_g \ket{g}$ on every site.


\emph{Non-Invertible Rep($G$) Symmetry}---It is well known that the TFIM and clock models \eqref{eq:TFIM} are self-dual, known as Kramers-Wannier duality, effectively exchanging the $X$ and $Z$ operators. 
The dual of \cref{eq:H_G} is (Appendix B)
\begin{equation}
    H_{\widetilde{G}} = -J \sum_{i} \sum_g \Xr{g}_i \Xl{g}_{i+1} - h \sum_i \sum_{\Gamma} d_\Gamma \Tr[Z_i^\Gamma] + \text{h.c.}
    \label{eq:H_RepG}
\end{equation}
Whereas \cref{eq:H_G} has a $G$ symmetry generated by the $X$ operators, \cref{eq:H_RepG} has a symmetry generated by the $Z$ operators, namely $\smash{\cramped{R_{\Gamma} = \Tr \prod_i Z_i^\Gamma}}$, where all neighboring indices of the $Z$'s are contracted. 
These are best expressed as matrix product operators (MPO), by treating the indices of the $Z$'s as the virtual legs of the MPO,
\begin{equation} \label{eq:syms}
    Z_{\alpha\beta}^\Gamma
    \equiv
    \begin{tikzpicture}[
        baseline={(0,0)}, 
        transform shape, 
        scale=0.9
        ]
        \draw (-0.5,0)--(0.5,0);
        \draw (0,-0.5)--(0,0.5);
        \node[mpo](t) at (0,0) {\small $Z^\Gamma$};
        \filldraw (-.5,0) circle (.5mm);
        \filldraw (+.5,0) circle (.5mm);
        \node[left]  at (-.5,0) {\small $\alpha$};
        \node[right] at (+.5,0) {\small $\beta$};
    \end{tikzpicture} 
    ,
    \quad
    R_\Gamma 
    =
    \,
    \begin{tikzpicture}[
        baseline={(0,-0.2)}, 
        transform shape, 
        scale=0.9
        ]
        \draw[rounded corners=5pt] (0,0)--(2.5,0)--(2.5,-0.6)--(-.5,-0.6)--(-.5,0)--(0,0);
        \draw (0,-0.5)--(0,0.5);
        \draw (2,-0.5)--(2,0.5);
        \node[mpo](t) at (0,0) {$Z^\Gamma$};
        \node[fill=white] at (1,0) {$\dots$};
        \node[mpo](t) at (2,0) {$Z^\Gamma$};
    \end{tikzpicture}
    \,.
\end{equation}
The symmetry operators $R_{\Gamma}$ multiply according to the representation algebra~\cite{paceSPTLSMTheoremsProjective2025,fechisinNoninvertibleSymmetryProtectedTopological2025,kobayashiProjectiveRepresentationsBogomolov2026, cuiperHouchesLectureNotes2026},
\begin{equation}
    R_{\Gamma_a} R_{\Gamma_b} = \sum_c N_{ab}^c R_{\Gamma_c}
    \,\,
    \Leftrightarrow
    \,\,
    \Gamma_a \otimes \Gamma_b = \bigoplus_c N_{ab}^c \Gamma_c,
    \label{eq:RepG_Algebra}
\end{equation}
where $N_{ab}^c$ are non-negative integers which count the number of times $\Gamma_c$ appears in the decomposition of $\Gamma_a \otimes \Gamma_b$ into irreps.
This  algebra is called Rep($G$), the dual of the $G$ symmetry.
For an Abelian group every irrep is 1-dimensional, the product of two irreps is another 1D irrep, and thus there is a unique irrep on the right hand side of \cref{eq:RepG_Algebra}, so the symmetry is invertible---the $R_\Gamma$ are unitary operators. 
This is Pontryagin duality---to wit, addition modulo $N$ (performed by $X$'s) is equivalent to multiplication of $N$'th roots of unity (performed by $Z$'s).
Non-Abelian groups have irreps with $d_{\Gamma}>1$, so some tensor products of irreps must be reducible, resulting in multiple terms on the right hand side of \cref{eq:RepG_Algebra}. 
This makes the Rep($G$) symmetry non-invertible when $G$ is non-Abelian---some of the $R_\Gamma$ have a non-trivial kernel~\footnote{
    A technical note regarding language: linear combinations of the $R_\Gamma$ operators form a fusion algebra, which may contain formal inverses, i.e. some linear combination $x_{\Gamma_a} = \sum_{b} c_b R_{\Gamma_b}$ such that $x_{\Gamma_a} R_{\Gamma_a} = 1$. The non-invertibility of the symmetry refers instead to treating the operators as forming a representation of the fusion category Rep($G$), for which \cref{eq:RepG_Algebra} encodes the fusion of simple objects.
}. 
The simplest way to see this is to consider the action of $R_{\Gamma}$ on a product state, which computes the character in irrep $\Gamma$ of the ordered product of group elements
\begin{equation}
    R_{\Gamma} \ket{g_1, \cdots, g_L} 
    = 
    \Tr[\Gamma^{g_1\cdots g_L}]
    \ket{g_1, \cdots, g_L}\,.
\end{equation}
If $d_\Gamma>1$ then at least one group element has zero character, so the $R_{\Gamma}$ annihilates some states and therefore cannot be inverted.

\emph{Non-Invertible Symmetry Breaking}---We now explore the spontaneous breaking of this non-invertible symmetry~\cite{bhardwajCategoricalLandauParadigm2024,bhardwajIllustratingCategoricalLandau2025,bhardwajLatticeModelsPhases2024,10.21468/SciPostPhys.17.4.115,Lootens2025-ng,kobayashiProjectiveRepresentationsBogomolov2026}.
To express the low-field symmetry broken ground states we introduce the dual of the $\ket{g}$ basis \cite{SM}, defined by 
\begin{equation}
    \ket{\Gamma_{\alpha\beta}} 
    = 
    \sqrt{\frac{d_{\Gamma}}{\vert G \vert}} 
    \sum_{ g \in G } 
    \Gamma_{\alpha\beta}^g 
    \ket{g}
    \equiv
    \begin{tikzpicture}[baseline={(0,0.0)}, transform shape, scale=0.9]
        \draw (-0.5, 0.0)--(0.5,0.0);
        \draw ( 0.0, 0.0)--(0.0,0.6);
        \node[mpo](t) at (0,0) {\small $\ket{\Gamma}$};
        \filldraw (-0.5,0) circle (.5mm);
        \filldraw (+0.5,0) circle (.5mm);
        \node[left]  at (-.5,0) {\small $\ket{\alpha}$};
        \node[right] at (+.5,0) {\small $\bra{\beta}$};
    \end{tikzpicture} 
    \,,
    \label{eq:dual_basis}
\end{equation}
where $\vert G \vert$ is the number of group elements. 
In the second equality we have expressed this as a matrix product state (MPS) tensor, with indices $\alpha$ and $\beta$ labeling virtual states transforming in irrep $\Gamma$ and its dual, respectively (Appendix C).
In this basis the $X$ operators are block-diagonalized, $\smash{\cramped{\Xl{g} \ket{\Gamma_{\alpha\beta}} = \Gamma^{g^{-1}}_{\alpha\gamma}\ket{\Gamma_{\gamma\beta}}}}$.
It follows that when $\cramped{h \to 0}$, the Rep($G$) symmetry-broken ground states of \cref{eq:H_RepG} are fully-contracted MPSs which are invariant under the action of the two-site $\smash{\Xr{} \Xl{}}$ term~\cite{SM}, 
\begin{equation}
    \ket{\bm{\Gamma}} 
    =
    \sum_{\{\alpha_i\}} \bigotimes_i \frac{1}{\sqrt{d_{\Gamma}}} \ket{\Gamma_{\alpha_i \alpha_{i+1}}}
    \,\,
    = 
    \,\,
    \begin{tikzpicture}[
        baseline={(0,-0.1)}, 
        transform shape, 
        scale=0.9
        ]
        \draw[rounded corners=4pt] (0,0)--(2.5,0)--(2.5,-0.5)--(-.5,-0.5)--(-.5,0)--(0,0);
        \draw (0,0)--(0,0.5);
        \draw (2,0)--(2,0.5);
        \node[mpo](t) at (0,0) {\small $\ket{\Gamma}$};
        \node[fill=white] at (1,0) {$\dots$};
        \node[mpo](t) at (2,0) {\small $\ket{\Gamma}$};
        \filldraw[fill=white] (+0.47, 0  ) circle (.6mm);
        \filldraw[fill=white] (+1.53, 0  ) circle (.6mm);
        \filldraw[fill=white] ( 1.00,-0.5) circle (.6mm);
    \end{tikzpicture}\,.
    \label{eq:Gamma_MPS_closed}
\end{equation}
The normalization of the MPS is naturally incorporated as a virtual bond insertion of the diagonal matrix $\Lambda_{\alpha\beta}\equiv\delta_{\alpha\beta}/\sqrt{d_{\Gamma}}$ formed by Schmidt eigenvalues for this MPS. It is indicated by the open circle. One can check that the MPS is expressed in its canonical form.
The $\cramped{\Xl{g\neq e}_i}$ and $\cramped{\Xr{g\neq e}_i}$ operators serve as local order parameters of spontaneous symmetry breaking. 
Indeed it is easy to show that  
$\smash{\bra{\bm \Gamma}\Xr{g\neq e}_i}\ket{\bm \Gamma} = \Tr[\Gamma^{g}]/d_{\Gamma}$~\cite{SM}. 
In contrast, deep in the paramagnetic regime ($\cramped{J\to 0}$) the ground state is given by the product state $\ket{\bm{e}}\equiv\bigotimes_i \ket{e}_i$, where $e$ is the group identity, 
and the order parameter vanishes, $\bra{\bm{e}}\Xr{g\neq e}_i\ket{\bm{e}} = 0$.

As an example, consider the smallest non-Abelian group---the dihedral group $D_3$, the symmetries of an equilateral triangle, generated by a 3-fold rotation $r$ and a reflection $s$, illustrated in \cref{fig:Ising-G-Qudits}.
Its character table is given in \cref{tab:D3}(a), with three conjugacy classes and three irreps---the 1D trivial irrep $A_1$, the 1D sign irrep $A_2$, and a 2D irrep $E$ acting as rotations in the plane, permuting the corners of the triangle.
Note that the 2D $E$ irrep has a zero character on the $[s]$ conjugacy class, reflecting the non-invertibility of Rep($D_3$) symmetry.
The Rep($D_3$) algebra is given in \cref{tab:D3}(b). 
We can see that $A_2$ generates an invertible $\mathbbm{Z}_2$ sub-symmetry, while $E$ generates the non-invertible part of the symmetry.

\begin{figure*}[t]
    \centering
    \includegraphics[width=\linewidth]{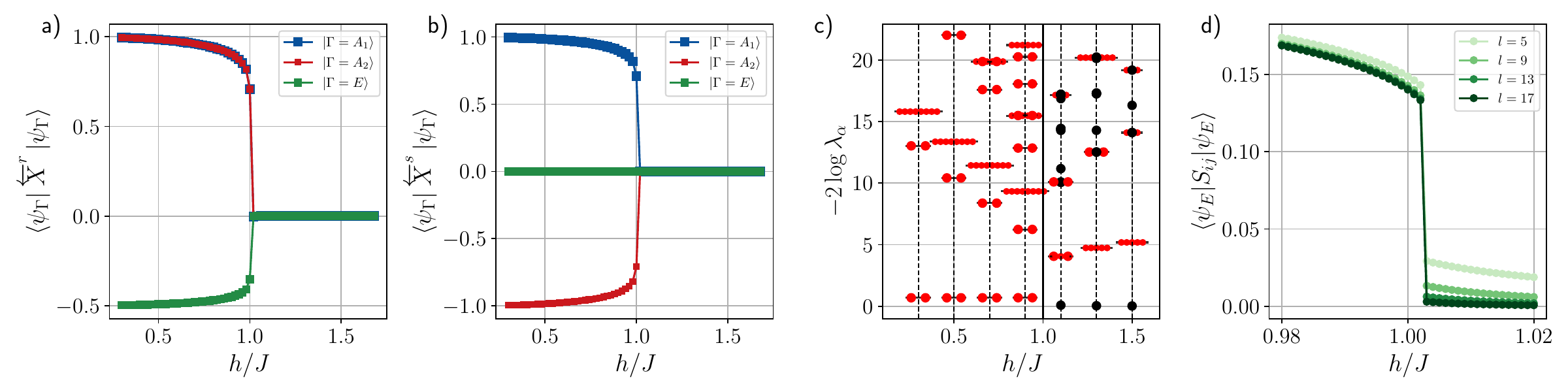}
    \caption{Spontaneously breaking a non-invertible Rep($G$) symmetry mixes local and non-local order. 
    Here we illustrate it for the group $D_3$ with the Hamiltonian \cref{eq:H_RepG} using iDMRG: 
    (a,b) Local order parameters $\smash{\langle \Xr{r}_i\rangle}$ and  $\smash{\langle \Xr{s}_i\rangle}$ in each symmetry-broken ground state; 
    (c) The entanglement spectrum under a bipartition of the ground state $\ket {\bm E}$. Non-degenerate Schmidt eigenvalues are colored black, while degenerate ones are colored red;
    (d) The string order parameter measured in the 2D irrep $\ket{\bm E}$ ground state for different string lengths. 
    All states are obtained by initializing in the exact $\ket{\bm \Gamma}$ ground state at $h=0$, then incrementally increasing $h/J$ and  numerically converging to the ground state in the same symmetry sector~\cite{Note2}.  
    }
    \label{fig:full_figure}
\end{figure*}

When $\cramped{h\to0}$ there are three symmetry broken ground states: two product states $\ket{\bm{A_1}}$,  $\ket{\bm{A_2}}$ and an MPS $\ket{\bm{E}}$ of bond-dimension 2. 
To study $\cramped{h > 0}$ we perform infinite density matrix renormalization group (iDMRG) calculations, initializing with one of the exact $h=0$ ground states \eqref{eq:Gamma_MPS_closed}, incrementally increasing $h$, and re-converging to the same symmetry-broken sector.
Figures \ref{fig:full_figure} (a,b) illustrate the expectation values for the local order parameters $\Xr{r}_i$ and $\Xr{s}_i$ as a function of $h$ for each ground state, showing a phase transition to the large-$h$ trivial phase~\footnote{
    The reader may notice that this phase transition occurs precisely at $\cramped{h=J}$. This is a consequence of the fine-tuned nature of \cref{eq:H_RepG}: it is dual to \cref{eq:H_G}, which can be written in the form $\cramped{H_G = -J\sum_i \delta(g_i, g_{i+1}) - h\sum_i \ket{+}_i\bra{+}_i}$, where we define the equal superposition state $\ket{+} = \vert G\vert^{-1/2}\sum_g \ket{g}$. This Hamiltonian is actually a $\vert G \vert$-state Potts model, which is self-dual~\cite{PhysRevB.24.218} with a  first-order transition if $\vert G \vert \geq 4$~\cite{igloi1983first}.
    Generic Hamiltonians \cref{apx_eq:H_G_generic,apx_eq:H_RepG_generic}, can exhibit continuous or discontinuous transitions depending on the parameters.
}
.

\begin{table}[t]
    \centering
    \begin{overpic}[width=\linewidth]{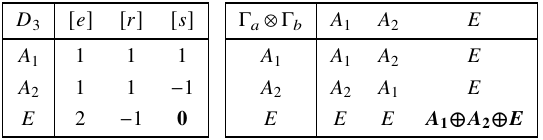}
        \put(20,-4){(a)}
        \put(70,-4){(b)}
    \end{overpic}
    \\[1ex]
    \caption{
    (a) The character table of the dihedral group $D_3$ (isomorphic to the permutation group $S_3$), where the characters are the traces of the irrep matrices $\Tr[\Gamma^g]$.
    The $E$ irrep has a zero character, which reflects the non-invertibility of Rep($D_3$) symmetry. 
    (b) The Rep($D_3$) algebra, where $A_1$ is the identity, $A_2$ generates a $\mathbbm{Z}_2$, and $E$ is a non-invertible generator due to $E\otimes E$ being reducible.
    }
    \label{tab:D3}
\end{table}

\emph{Ground State Entanglement}---So far non-invertible Rep($G$) symmetry breaking does not look so different from ordinary $G$ symmetry breaking, but let us juxtapose them more closely.
Symmetry broken ground states of the $G$-symmetric Hamiltonian \cref{eq:H_G} are labeled by a group element, and any two ground states are related by the symmetry action, $U_g\ket{\bm{h}} = \ket{\bm{gh}}$.
This is not the case in the non-invertible case: instead we have 
$R_{\Gamma_a} \ket{\bm{\Gamma}_b} = \sum_c N_{ab}^c \ket{\bm{\Gamma}_c}$.
In particular, the action of the symmetry on a ground state can create a linear superposition of macroscopically distinct ground states, a cat state, which never happens for invertible symmetries.

Notice that the bond dimension of the ground state MPS in \cref{eq:Gamma_MPS_closed} is equal to the dimension of the irrep $\Gamma$, indicating that \emph{different ground states have inequivalent entanglement structures}---1D irreps are unentangled product states, while higher-dimensional ones carry non-trivial entanglement. 
Such entanglement structure is a hallmark of SPT order, for which it is reflected in (i) the presence of zero modes at the boundary of an open system, (ii) exact degeneracy of the entanglement spectrum, and (iii) the existence of a non-local string order parameter. 
We will now show that all three of these SPT markers are present for Rep($G$) symmetry broken ground states with $d_{\Gamma}>1$.

First, whereas on a closed chain the number of ground states is equal to the number of irreps, on an open chain the ends of the MPS tensors \cref{eq:Gamma_MPS_closed} are no longer contracted. For a given irrep $\Gamma$, the ground states carry free indices at the ends
\begin{equation}
    \ket{\bm{\Gamma}_{\alpha\beta}} 
    \,\,
    = 
    \,\,
    \begin{tikzpicture}[baseline={(0,-0.1)}, transform shape, scale=0.9]
        \draw (-.5,0)--(2.5,0);
        \draw (0,0)--(0,0.5);
        \draw (2,0)--(2,0.5);
        \node[mpo](t) at (0,0) {\small $\ket{\Gamma}$};
        \node[fill=white] at (1,0) {$\dots$};
        \node[mpo](t) at (2,0) {\small $\ket{\Gamma}$};
        \filldraw (-0.5,0) circle (.5mm);
        \filldraw (+2.5,0) circle (.5mm);
        \node[left]  at (-0.5,0) {\small $\ket \alpha$};
        \node[right] at (+2.5,0) {\small $\bra \beta$};
        \filldraw[fill=white] (+0.47,0) circle (.6mm);
        \filldraw[fill=white] (+1.53,0) circle (.6mm);
    \end{tikzpicture}\,.
    \label{eq:Gamma_MPS_open}
\end{equation}
The number of distinct ground states on an open chain is therefore $\sum_{\Gamma}d_\Gamma^2 = \vert G \vert$, which we verified numerically with open-chain DMRG calculations.
For each irrep with $d_{\Gamma}>1$ the symmetry broken ground state has localized gapless edge modes.
Since $\smash{\cramped{\Xl{g}_L}}$ at the left end of the chains and $\smash{\cramped{\Xr{g}_R}}$ at the right end of the chain commute with the Hamiltonian, but do not commute with the Rep($G$) symmetry, they are strong zero modes~\cite{aliceaTopologicalPhasesParafermions2016}.
By acting on an energy eigenstate, these local operators generate a non-equivalent state of the same energy but with a different internal state of the non-contracted external leg.
In summary, ground states $\ket{\bm{\Gamma}_{\alpha\beta}}$ span the Hilbert space of a single $G$-qudit. 
In contrast to ordinary symmetry breaking, part of it is fractionalized between the two ends of the chain. 

Second, we obtain the entanglement spectrum (the spectrum of the reduced density matrix of a bipartition) using iDMRG for each ground state as $h$ is varied. \Cref{fig:full_figure}(c) shows the Schmidt eigenvalues for the $\ket{\bm{E}}$ ground state of the Rep($D_3$)-symmetric Hamiltonian, which exhibit exact double degeneracies of every eigenvalue throughout the symmetry broken phase, compared to singlet eigenvalues in the trivial phase.
Beyond $D_3$, we find that the degeneracy is robust---a $\ket{\bm{\Gamma}}$ ground state has a $d_{\Gamma}$-degenerate entanglement spectrum within the SSB phase (Appendix D).
There we have also checked that these entanglement degeneracies are robust to generic Rep($G$)-preserving perturbations of the Hamiltonian which break any additional global symmetries.

The existence of entangled SSB states is an intrinsic property of the non-invertible symmetry originating from the structure of its symmetry multiplets, which mix local and non-local operators~\cite{bhardwajLatticeModelsPhases2024,bhardwajGappedPhasesNoninvertible2025}. 
Consider first an ordinary $G$ symmetry: local charged operators form irreducible multiplets which transform as 
$\smash{\cramped{U_g \mathcal{O}_{\Gamma}^\alpha = (\Gamma^g_{\alpha\beta}\mathcal{O}_{\Gamma}^{\raisemath{-1pt}{\smash{\beta}}}) U_g}}$.
In contrast, for the order parameters of $\Rep(D_3)$ symmetry, we can derive the operator equality
\begin{equation}
    R_E \,\Xl{r}_j 
    = 
    \left[
    \Re(\omega)\, 
    \Xl{r}_j
    + 
    i \,\Im(\omega)\, 
    \Big(\prod_{k< j} Z_k^{A_2}\Big)\,
    \Xl{r}_j
    \right]
    R_E,
    \label{eq:E_r_multiplet}
\end{equation}
where $\omega = \exp(i2\pi /3)$ (Appendix F). 
This demonstrates that the local operator $\Xl{r}$ forms an irreducible multiplet together with a non-local string operator, a remarkable property ensured by the non-invertible Rep($G$) symmetry.
In particular in the SSB phase this implies that if a local order parameter is non-zero then a string order parameter from the same multiplet must also be non-zero.
Starting from \eqref{eq:E_r_multiplet}, in~\cite{SM}  we derive relations between two-point correlation functions of local charged operators and string order parameters. 
Using iDMRG, we observe in \cref{fig:full_figure}(d) that the string order parameter $S_{ij}=\cramped{\smash{\Xr{r}_i \prod_{i < k < j} Z_k^{\raisemath{-1pt}{\smash{A_2}}} \Xl{r}_j}}$ has a vacuum expectation value in the $\ket{\bm{E}}$ ground state.

\emph{Anyonic Domain Walls}---The mixing of local and non-local order and the presence of non-trivial entanglement in the non-invertible symmetry broken phase is reflected in the nature of the domain wall  excitations.
Letting $\Gamma_0$ denote the trivial representation, the state $\ket{\bm{\Gamma}_0}$ plays a special role---any ground state can be created as $\ket{\bm{\Gamma}} = R_{\Gamma} \ket{\bm{\Gamma}_0}$ since $\Gamma_0$ acts as the identity element of the Rep($G$) algebra.
Acting on $\ket{\bm{\Gamma}_0}$ with a truncated $R_{\Gamma_a}$ MPO creates a $\Gamma_a$ domain inside the $\Gamma_0$ ground state, with a $\Gamma_a$ ($\Gamma_{\bar a}$) domain wall at the left (right) end, where $\bar{a}$ denotes the dual representation (Appendix C).
Since the truncated MPO contains an uncontracted virtual bond, each domain wall carries an internal $d_{\Gamma_a}\!$-dimensional Hilbert space.  
These quasiparticles fuse according to the tensor product of their corresponding irreps
\begin{equation}
    \ket{\psi_{ab}} = \!\begin{tikzpicture}[
        baseline={(0,-0.1)}, 
        transform shape, 
        scale=0.9
        ]
        \node[left]  at (-2.7,0) {$\dots$};
        \node[right] at ( 2.8,0) {$\dots$};
        \draw (-2.8,0)--(-0.4,0);
        \draw ( 0.4,0)--( 2.8,0);
        \draw (-2.3,0)--(-2.3,0.7);
        \draw (-1.2,0)--(-1.2,0.7);
        \draw ( 1.2,0)--( 1.2,0.7);
        \draw ( 2.3,0)--( 2.3,0.7);
        \node[mpo](t) at (-2.3,0) {\small $\ket{\Gamma_a}$};
        \node[mpo](t) at (-1.2,0) {\small $\ket{\Gamma_a}$};   
        \node[mpo](t) at ( 1.2,0) {\small $\ket{\Gamma_b}$};
        \node[mpo](t) at ( 2.3,0) {\small $\ket{\Gamma_b}$};
        \filldraw (-0.4,0) circle (.5mm);
        \filldraw ( 0.4,0) circle (.5mm);
        \node at (-.4,.3) {\small $\ket{\psi_{\bar a}}$};
        \node at ( .4,.3) {\small $\ket{\psi_b}$};
        \filldraw[fill=white] (-1.75,0) circle (.6mm);
        \filldraw[fill=white] (+1.75,0) circle (.6mm);
    \end{tikzpicture}
    \label{eq:psi_ab_unfused}
\end{equation}
where $\ket{\psi_b}$ and $\cramped{\ket{\psi_{\bar a}}\equiv \bra{\psi_a}}$ denote the internal states of the quasiparticles, with virtual Hilbert space dimensions $d_{\Gamma_{\bar a}}$ and $d_{\Gamma_b}$ respectively. 
We can make a unitary change of basis for the joint internal state of the two quasiparticles, $\cramped{\ket{\psi_{\bar a}}\otimes \ket{\psi_b}}$, decomposing into different irreducible fusion channels using the Clebsch-Gordan decomposition of the tensor product $\cramped{\Gamma_{\bar a}\otimes \Gamma_b = \bigoplus_c N_{\bar{a}b}^c \Gamma_c}$ (Appendix E),
\begin{equation}
    \ket{\Gamma_{c_n},\gamma} = \sum_{\alpha,\beta} 
    [\mathscr{C}_{\bar{a}b}^{c_n}]_{\bar{\alpha}\beta}^\gamma
    \,\,
    \ket{\Gamma_{\bar a},\bar \alpha} \otimes \ket{\Gamma_b ,\beta},
    \label{eq:Clebsch_Gordan}
\end{equation}
where $n=1\ldots N_{\bar{a}b}^c$ labels the copies of $\Gamma_c$ in the direct sum, and the greek indices label the basis states of each irrep.
The Clebsch-Gordan coefficients may be viewed as a unitary matrix $\mathscr{C}_{\bar{a}b}$ with rows indexed by $(c_n,\gamma)$ and columns labeled by $(\bar{\alpha},\beta)$.
Expanding $\smash{\cramped{\ket{\psi_{\bar a}}=\sum_{\bar \alpha=1}^{d_{\Gamma_{\bar a}}}\psi_{\bar a}^{\bar\alpha}\ket{\Gamma_{\bar a},\bar\alpha}}}$ and similarly for $\ket{\psi_b}$, the two domain wall state \cref{eq:psi_ab_unfused} can then be decomposed as
\begin{equation}
    \ket{\psi_{ab}} 
    = 
    \sum_{c}
    \sum_{n=1}^{N_{ab}^c} 
    \!
    \begin{tikzpicture}[
        baseline={(0,0)}, 
        transform shape, 
        scale=0.85
        ]
        \node[left]  at (-2.4,0) {$\dots$};
        \node[right] at ( 2.4,0) {$\dots$};
        \draw (-2.45,0)--(2.45,0);
        \draw (-1.9,0)--(-1.9,0.9);
        \draw (-0.9,0)--(-0.9,0.9);
        \draw ( 0.0,0)--( 0.0,0.5);
        \draw ( 0.9,0)--( 0.9,0.9);
        \draw ( 1.9,0)--( 1.9,0.9);
        \node[mpo](t) at (-1.9,0) {\small $\ket{\Gamma_a}$};
        \node[mpo](t) at (-0.9,0) {\small $\ket{\Gamma_a}$};   
        \node[mpo](t) at ( 0.9,0) {\small $\ket{\Gamma_b}$};
        \node[mpo](t) at ( 1.9,0) {\small $\ket{\Gamma_b}$};
        \draw[fill=white, draw=black] (0,0) circle (.35cm) node {\small $\mathscr{C}_{c_n}^{\bar{a}b}$};
        \filldraw[fill=white] (-1.4,0) circle (.6mm);
        \filldraw[fill=white] (+1.4,0) circle (.6mm);
        \filldraw (0,0.5) circle (.5mm);
        \node[above] at (0,0.5) {\small $\ket{\psi_c}$};
    \end{tikzpicture}
    \label{eq:psi_ab_fused}
\end{equation}
where the virtual bond insertions are $\cramped{d_{\Gamma_{\bar a}}\!\times d_{\Gamma_b}}$ matrices constructed from the Clebsch-Gordan coefficients which encode the internal state of a $\Gamma_c$ quasiparticle,
\begin{equation}
    \begin{tikzpicture}[
        baseline={(0,-0.1)}, 
        transform shape, 
        scale=0.9
        ]
        \draw (-0.6,0)--(0.6,0.0);
        \draw ( 0.0,0)--(0.0,0.55);
        \draw[fill=white, draw=black] (0,0) circle (.35cm) node {\small ${\mathscr{C}_{c_n}^{\bar{a}b}}$};
        \node[left]  at (-.6,0) {\small $\braket{\bar\alpha\vert\psi_{\bar a}}$};
        \node[right] at (+.6,0) {\small $\braket{\beta\vert\psi_b}$};
        \filldraw (-.6,0) circle (.5mm);
        \filldraw (+.6,0) circle (.5mm);
        \filldraw (0,0.55) circle (.5mm);
        \node[above] at (0,0.6) {\small $\ket{\psi_{c_n}}$};
    \end{tikzpicture}
    \equiv 
    \sum_{\gamma}
    [\mathscr{C}^{{\bar a} b}_{c_n}]^{\bar\alpha\beta}_\gamma
    \,\psi_{\bar a}^{\bar\alpha} \, \psi_b^\beta
    \,\,
    \ket{\Gamma_c,\gamma}
    ,
    \label{eq:CG_MPS_defect}
\end{equation}
where $\mathscr{C}^{{\bar a} b}$ denotes the inverse of $\mathscr{C}_{{\bar a} b}$.
\Cref{eq:psi_ab_fused} describes a superposition of states with a single $\Gamma_c$ quasiparticle at the interface between a $\Gamma_a$ and $\Gamma_b$ domain, whose internal state depends on the initial states $\ket{\psi_{\bar a}}$ and $\ket{\psi_b}$.
Thus colliding the $\Gamma_a$ and $\Gamma_b$ quasiparticles gives rise to different possible fusion channels, which is a defining property of non-Abelian anyons \cite{nayakNonAbelianAnyonsTopological2008,simonTopologicalQuantum2023}.
Notably, if $\cramped{\Gamma_c \in \Gamma_{\bar{a}}\otimes \Gamma_a}$, then a $\Gamma_c$ quasiparticle can propagate within a $\Gamma_a$ domain.

We have seen that Rep($G$) SSB ground states, by virtue of being labeled by the objects of a fusion category, behave differently from the familiar invertible SSB and share many properties with SPTs. 
Notably, a static junction between distinct SPTs protected by invertible symmetries necessarily carries zero modes.
Despite not being SPTs in the ordinary sense, certain domain wall interfaces between different Rep($G$) SSB ground states carry zero modes, and so behave like dynamical SPT junctions.

Using \cref{eq:psi_ab_fused} we can also understand the origin of the entanglement spectrum degeneracies.
For an SPT protected by an invertible group symmetry, the action of a truncated symmetry operator $U_g$ on the ground state inserts a symmetry defect $V_g$ on the virtual leg at its ends; for Abelian symmetries, if different defects do not commute, $[V_g,V_{g'}]\neq 0$, the representation must be projective, indicating symmetry fractionalization and a protected ES degeneracy \cite{pollmannEntanglementSpectrumTopological2010}. 
A similar argument can be applied to the ES degeneracies of Rep($G$) SSB states: the defects \eqref{eq:CG_MPS_defect} are inserted at the ends of truncated symmetry operators; focusing on the defects $\mathscr{C}^{\bar{a}a}_{c_n}$ which can appear in a single $\ket{\bm{\Gamma}_a}$ ground state, the failure of these defects to commute (despite the commutation of the $R_{\Gamma}$ generators of Rep($G$) symmetry) implies degeneracy of the entanglement spectrum. 
This happens provided the defects commute with the matrix of Schmidt eigenvalues.
As shown in Appendix F, in the Rep($D_3$) case the $\smash{\mathscr{C}^{\overline EE}_{A_2}}$ defect in the $\ket{\bm{E}}$ ground state is $\sigma^z$, while the two $\smash{\mathscr{C}^{\overline EE}_{E}}$ defects are $\sigma^{\pm}$, implying two-fold degeneracy of the entanglement spectrum of this symmetry-broken ground state.

\emph{Outlook}---Our work connects closely to discrete lattice gauge theories such as 2D quantum double models~\cite{kitaevFaulttolerantQuantumComputation2003}---non-Abelian generalizations of the $\mathbbm{Z}_2$ toric code---which are naturally built from $G$-qudits.
In this context, the Ising-type terms in the Hamiltonian \cref{eq:H_RepG} may be viewed as 1D equivalents of the ``star'' operators which measure the local electric charge. 
Taking this perspective, the Rep($G$) SSB ground states are those with zero charge.
The symmetry operators $R_\Gamma$ are the Wilson line operators of the gauge theory---creation operators of electric field lines.
The trivial irrep corresponds to the zero-field state, and each state $\ket{\bm{\Gamma}}=R_{\Gamma} \ket{\bm{\Gamma}_{0}}$ corresponds to a single non-Abelian electric field string running through the system.
The domain walls are non-Abelian electric charges.
Alternatively, our 1D model may be viewed as describing the gapped edge of a quantum double model~\cite{albertSpinChainsDefects2021}, closely related to the so-called SymTFT construction~\cite{bhardwajCategoricalLandauParadigm2024,jiCategoricalSymmetry2020,paceSPTLSMTheoremsProjective2025,bhardwajGappedPhasesNoninvertible2025}. 
As with quantum double models, our model can be further generalized from finite groups to Hopf algebras~\cite{jiaGeneralizedClusterStates2024,jiaWeakHopfNoninvertible2026,BuerschaperHierarchyTopologicalTensorNetwork2013,Inamura2022-yv,luGeneralizedKramersWannierSelfDuality2026}. 

Significant effort has been put towards realizing 2D non-Abelian topological order for anyon quantum computation~\cite{nayakNonAbelianAnyonsTopological2008,bylesDemonstrationMagicState2025, loUniversalQuantumComputation2025}. 
A natural intermediate step towards that goal is to realize 1D $G$-qudit models with non-invertible Rep($G$) symmetry. 
An implementation using neutral atoms trapped in optical tweezer arrays has been proposed in Ref.~\cite{249m-m8wq}.
An alternative promising direction is to prepare the fixed-point Rep($G$) symmetry-broken ground states, \cref{eq:Gamma_MPS_open}, and their cat superpositions~\cite{kobayashiProjectiveRepresentationsBogomolov2026} using existing quantum hardware based on trapped ions of calcium atoms \cite{hrmoNativeQuditEntanglement2023} which allows to implement qudits with dimension up to $\cramped{d=7}$. 
This can be done by constructing sequential circuits made of two-qudit entangling controlled-$X$ gates~\cite{brellGeneralizedClusterStates2015,fechisinNoninvertibleSymmetryProtectedTopological2025,kobayashiProjectiveRepresentationsBogomolov2026}. Moreover, Ref.~\cite{edmundsSymmetryProtectedTopologicalHaldane2025} demonstrates the ability to prepare an SPT state, manipulate its edge modes, and measure local and string order parameters on existing quantum trapped-ion hardware. These methods can be directly transferred to the preparation, manipulation and diagnostics of the Rep($G$) SSB states investigated here.

\emph{Acknowledgments}---We thank Z. Komargodski and B. Rayhaun for fruitful discussions, and P. Sala for helpful comments on the manuscript. The theoretical contribution by A.H.N. was supported by the Department of Energy under the Basic Energy Sciences award no. DE-SC0025047.
This work was in part supported by the Deutsche Forschungsgemeinschaft under the cluster of excellence ct.qmat (EXC-2147, project number 390858490). S.M. is supported by Vetenskapsr{\aa}det (grant number 2021-03685) and Nordita.


%

 \clearpage

 \onecolumngrid
 \vspace{\columnsep}
 \begin{center}
 \textbf{End Matter}
 \end{center}
 \vspace{\columnsep}
 \twocolumngrid

\appendix{}

\renewcommand{\theequation}{A.\arabic{equation}} 
\setcounter{equation}{0}
\textit{Appendix A: General Form of Hamiltonian}---In the main text we presented only a special fine-tuned version of the nearest-neighbor generalized TFIM Hamiltonians for $G$ symmetry, \cref{eq:H_G}, and Rep($G$) symmetry, \cref{eq:H_RepG}. The most general $G$-symmetric Hamiltonian is
\begin{equation}
    H_G 
    = 
    -
    \sum_{i} 
    \sum_{\Gamma}
    \sum_{\alpha,\beta=1}^{d_{\Gamma}}
    J^{\Gamma}_{\alpha\beta}
    [
    Z_{i}^{\overline \Gamma}\cdot Z_{i+1}^{\Gamma}
    ]_{\alpha\beta} 
    - 
    \sum_i 
    \sum_{g \in G} 
    h_g 
    \Xr{g} + \text{h.c.}
    \label{apx_eq:H_G_generic}
\end{equation}
with the $G$ symmetry generated by $\cramped{U_g = \prod_i \Xl{g}_i}$.
So long as all $\cramped{J^{\Gamma}_{\alpha\beta}} > 0$, the ground state as $h_g \to 0$ completely breaks the $G$ symmetry, with one ground state for each group element $g$. 
The most general Rep($G$)-symmetric Hamiltonian is
\begin{equation}
    H_{\Rep(G)}
    =
    -
    \sum_i 
    \sum_{g\in G}
    J_g 
    \Xr{g}_i 
    \Xl{g}_{i+1}
    -
    \sum_i 
    \sum_{\Gamma}
    \sum_{\mathclap{\alpha,\beta=1}}^{d_\Gamma}
    h^{\Gamma}_{\alpha\beta}
    [Z_i^\Gamma]_{\alpha\beta}+ \text{h.c.}
    \label{apx_eq:H_RepG_generic}
\end{equation}
this completely breaks the Rep($G$) symmetry when $h^{\Gamma}_{\alpha\beta} \to 0$ so long as all $\cramped{J_g > 0}$, with one ground state for each irrep $\Gamma$ on a closed chain, \cref{eq:Gamma_MPS_closed}.

\renewcommand{\theequation}{B.\arabic{equation}} 
\setcounter{equation}{0}
\textit{Appendix B: Gauging Duality}---Here, we review how gauging relates a $G$-symmetric model to a dual Rep($G$)-symmetric model, \cref{apx_eq:H_G_generic} and \cref{apx_eq:H_RepG_generic}. 
For Abelian groups these are isomorphic, so it is a self-duality called Kramers-Wannier duality, while for non-Abelian groups it relates two distinct models. 
Here we sketch how this works: starting from the $G$-symmetric \cref{apx_eq:H_G_generic}, we fully gauge the $G$ symmetry, meaning that we minimally couple the $ZZ$ term to a $G$-valued gauge field on the links and enforce the Gauss law. 
The Hilbert space of a $G$ gauge field is nothing but a $G$ qudit, and we denote the corresponding link operators with $\mathcal{Z}$ and $\mathcal{X}$. 
The minimally-coupled gauged Hamiltonian is
\begin{equation}
    H_{G}
    \!
    \to
    -\sum_i \sum_{\Gamma,\alpha\beta} J^{\Gamma}_{\alpha\beta}
    [
    Z^{\overline \Gamma}_{i} 
    \!\cdot
    \z^\Gamma_{i+\frac{1}{2}} 
    \!\!\cdot
    Z^{\Gamma}_{i+1}
    ]_{\alpha\beta}
    -\sum_i \sum_g h_g \Xr{g}_i
    + \text{h.c.}
\end{equation}
The operators
$
    \smash{
    \cramped{
    G_i^{g} 
    \equiv 
    \xr{g}_{\raisemath{-1pt}{\smash{i-\scriptscriptstyle{\frac{1}{2}}}}}
    \Xl{g}_i \,
    \xl{g}_{\raisemath{-1pt}{\smash{i+\scriptscriptstyle{\frac{1}{2}}}}}
    }
    }
$
perform gauge transformations, and we enforce the Gauss constraint $\smash{\cramped{G_i^g\overset{\raisebox{-.1\height}{$\scriptscriptstyle !$}}{=} 1}}$.
This constraint can be resolved by fixing to unitary gauge, where every original $G$-qudit on the sites is rotated to the state $\ket{e}$. Acting with $\Xr{g}_i$ changes this state to $\ket{g^{-1}}$, which is undone by a $\smash{G_i^{g}}$ gauge transformation, meaning that in the gauge-fixed Hilbert space $\Xr{g}_i$ acts as 
$
 \smash{
 \xr{g}_{\raisemath{-1pt}{\smash{i-\frac{1}{2}}}} 
 \xl{g}_{\raisemath{-1pt}{\smash{i+\frac{1}{2}}}}
 }
$.
Meanwhile, the $[Z_i^\Gamma]_{\alpha\beta}$ acting on this gauge-fixed state produce Kronecker deltas $\delta_{\alpha\beta}$.
Thus the gauge-fixed Hamiltonian is
\begin{equation}
    H_{\Rep(G)}
    \!
    =
    \!
    -
    \!
    \sum_i \sum_{\Gamma,\alpha\beta} J^{\Gamma}_{\alpha\beta} [\z^\Gamma_{i+\frac{1}{2}}]_{\alpha\beta}
    -\sum_i \sum_g h_g 
    \xr{g}_{\raisemath{-1pt}{\smash{i-\frac{1}{2}}}} 
    \xl{g}_{\raisemath{-1pt}{\smash{i+\frac{1}{2}}}}+ \text{h.c.}
\end{equation}
which is \cref{apx_eq:H_RepG_generic} after relabeling $h \leftrightarrow J$, and translating by half a site.

\renewcommand{\theequation}{C.\arabic{equation}} 
\setcounter{equation}{0}

\textit{Appendix C: Peter-Weyl Decomposition of $G$-qudit Hilbert Space}---Mathematically, the Peter-Weyl theorem applied to finite groups states that the Hilbert space of a single $G$-qudit decomposes as~\cite{marianiHamiltoniansGaugeinvariantHilbert2023}
\begin{equation}
    \mathscr{H} \cong \bigoplus_{\Gamma} V_{\Gamma} \otimes {V}_{\Gamma}^*,
\end{equation}
where $V_{\Gamma}$ is a $d_{\Gamma}$-dimensional vector space and $\smash{V_{\Gamma}^*}$ is its dual. 
The basis $\ket{\Gamma_{\alpha\beta}}$, \cref{eq:dual_basis}, precisely encodes this decomposition, i.e. we may write $\smash{\cramped{\ket{\Gamma_{\alpha\beta}} \equiv \ket{\Gamma_{\alpha}}\otimes \bra{{\Gamma}_{\beta}^*}}}$, where $\ket{\Gamma_\alpha}$ form an orthonormal basis for $V_{\Gamma}$ and $\smash{\bra{\Gamma_\beta^*}}$ are the dual basis vectors.
The operators $\smash{\Xl{g}}$ act on $\smash{V_{\Gamma}}$ and the operators $\smash{\Xr{g}}$ act on $\smash{V_{\Gamma}^*}$.
When considering the states $\ket{\Gamma_{\alpha\beta}}$ as MPS tensors, the left index labels states $\ket{\Gamma_\alpha}$ in the virtual Hilbert space $V_{\Gamma}$, and the right index labels states $\bra{\Gamma_\beta^*}$ in the dual space $V_{\Gamma}^*$.
Throughout this text, we denote the dual representation by $\Gamma_a^*\equiv \Gamma_{\bar a}$.

\renewcommand{\theequation}{D.\arabic{equation}} 
\setcounter{equation}{0}

\begin{figure}[b]
    \centering
    \includegraphics[width=\linewidth]{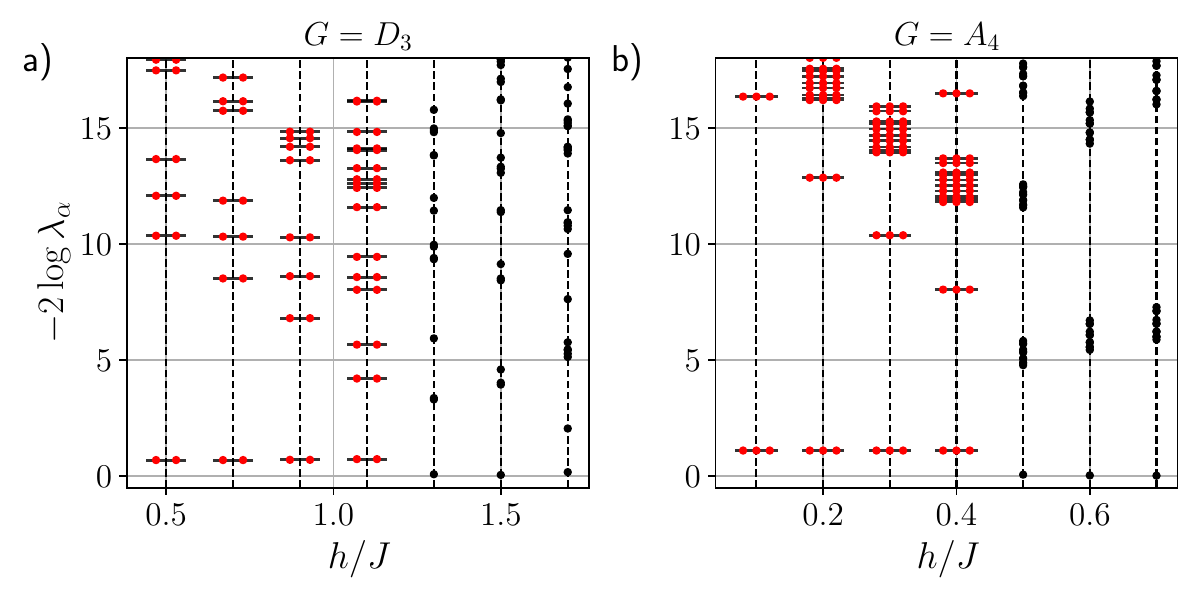}
    \caption{Entanglement spectra for $\Rep(G)$ symmetric Hamiltonians with generic complex $J_g$ couplings as a function of the ``transverse field'' $h$ for the groups $D_3$ (a) and $A_4$ (b). Both groups have exactly one representation of dimension $d_{\Gamma}>1$, with $d_{\Gamma}=2$ and $3$, respectively. For small values of $h$ the symmetry broken ground states corresponding to such representations exhibit $d_{\Gamma}$-fold degeneracy of the entanglement spectrum.}
    \label{fig:ent_spec_sup}
\end{figure}

\textit{Appendix D: Entanglement spectrum degeneracies for generic $\Rep(G)$ symmetric Hamiltonians}---In the main text we focused our analysis on the simplest case of $\Rep(G)$-symmetric Hamiltonians with couplings which are real and uniform. Here, we show how different generic choices do not affect our conclusions regarding the degeneracies of the entanglement spectrum, which is therefore to be attributed entirely to the $\Rep(G)$ invariance and not to other accidental symmetries. In \cref{fig:ent_spec_sup} we present the entanglement spectra for the group $D_3$, addressed already in the main text, and for the alternating group $A_4$.  The latter is a non-Abelian group with $12$ elements, three one-dimensional irreps and one three-dimensional irrep. In contrast to \cref{fig:full_figure}, we now consider group-element dependent couplings $J_g=J+\delta J_g$, where the deviations $\delta J_g$ are small enough not to modify the symmetry breaking pattern (e.g. they do not introduce ``antiferromagnetic'' order), but break the symmetry which rotates between the different group elements.  $\delta J_g$ are chosen to be complex, thus breaking time reversal. In the ordered phase, we target the ground state corresponding to the high-dimensional irreps by choosing an appropriate initial state for the iDMRG simulations. We observe that up until the phase transition point the whole entanglement spectrum of the system exhibits exact $d_{\Gamma}$-fold degeneracies.

\renewcommand{\theequation}{E.\arabic{equation}} 
\setcounter{equation}{0}

\textit{Appendix E: Clebsch-Gordan Coefficients, Fusion, and Operator Multiplets}---The Clebsch-Gordan coefficients $\mathscr{C}_{ab}^{c_n}$ of the group $G$, \cref{eq:Clebsch_Gordan}, play a key role in the fusion of defects and in determining the symmetry multiplets of operators. 
This is because they encode the intertwiners between representations, i.e. linear maps $\cramped{\phi_{ab}^{c_n}:V_{\Gamma_a}\otimes V_{\Gamma_b} \to V_{\Gamma_{c_n}}},$
\begin{equation}
    \phi_{ab}^{c_n}= \sum_{\alpha,\beta,\gamma} 
    [\mathscr{C}_{ab}^{c_n}]_{\alpha\beta}^\gamma \, 
    \ket{\Gamma_{c_n},\gamma}
    \otimes \bra{\Gamma_a,\alpha} \otimes \bra{\Gamma_b,\beta}\,.
\end{equation}
where $\ket{\Gamma_a,\alpha}$ is a basis for $V_{\Gamma_a}$. Note that for dual irreps $\Gamma_{\bar{a}}$ acting on $V_{\Gamma_a}^*$, we have the dual basis $\cramped{\ket{\Gamma_{\bar{a}},\bar{\alpha}}\equiv \bra{\Gamma_a,\alpha}}$.
We can write the Clebsch-Gordan as a trivalent junction,
\begin{equation}
    [\mathscr{C}_{ab}^{c_n}]_{\alpha\beta}^\gamma 
    \,\,
    \equiv
    \,
    \begin{tikzpicture}[baseline={(0,0)}]
        \draw (0,0)--( 0.0, 0.6);
        \draw (0,0)--(-0.6, 0.0);
        \draw (0,0)--( 0.6, 0.0);
        \filldraw ( 0.0, 0.6) circle (.5mm);
        \filldraw (-0.6,-0.0) circle (.5mm);
        \filldraw ( 0.6,-0.0) circle (.5mm);
        \draw[fill=white, draw=black] (0,0) circle (.35cm) node {\small $\mathscr{C}_{ab}^{c_n}$};
        \node[left]  at (-0.6, 0.0) {\small $\alpha$};
        \node[right] at ( 0.6, 0.0) {\small $\beta$};
        \node[above] at ( 0.0, 0.65) {\small $\gamma$};
    \end{tikzpicture}
    \,
    .
\end{equation}
These are the matrices that appear on the MPS virtual legs when fusing an $a$ and $b$ domain walls, \cref{eq:CG_MPS_defect}.
Since they block-diagonalize the tensor products of representation matrices, they do the same when multiplying $Z$ operators,
\begin{equation}
    Z^{\Gamma_a}_{\alpha\alpha'} Z^{\Gamma_b}_{\beta\beta'} = \sum_{\Gamma_c \in \Gamma_a \otimes \Gamma_b} \sum_{n=1}^{N_{ab}^c} 
    \sum_{\gamma,\gamma'=1}^{d_{\Gamma_c}}
    [\mathscr{C}_{ab}^{c_n}]_{\alpha\beta}^{\gamma} 
    Z_{\gamma\gamma'}^{\Gamma_c} 
    [\mathscr{C}^{ab}_{c_n}]^{\alpha'\beta'}_{\gamma'},
\end{equation}
where $\mathscr{C}^{ab}$ is the inverse of the unitary Clebsch-Gordan matrix $\mathscr{C}_{ab}$ with rows indexed by $(c_n,\gamma)$ and columns indexed by $(\alpha,\beta)$. 
In particular, this tells us how to fuse truncated symmetry operators,
\begin{equation}
    \begin{tikzpicture}[
        baseline={(0,0.5)}, 
        transform shape, 
        scale=1
        ]
        \draw (0.4,0.0)--(2.6,0.0);
        \draw (1.4,0.8)--(2.6,0.8);
        \draw ( 1.0,-0.5)--( 1.0,1.3);
        \draw ( 2.0,-0.5)--( 2.0,1.3);
        \node[mpo](t) at (2.0,0.8) {\small $Z^{b}$};
        \node[mpo](t) at (1.0,0.0) {\small $Z^{a}$};
        \node[mpo](t) at (2.0,0.0) {\small $Z^{a}$};
        \filldraw (1.4,0.8) circle (.5mm);
        \node[above] at (1.4,.87) {\small $\beta$};
    \end{tikzpicture}
    =
    \sum_{c \in a\otimes b}
    \sum_{n=1}^{N_{ab}^c} 
    \,
    \begin{tikzpicture}[
        baseline={(0,-0.1)}, 
        transform shape, 
        scale=1
        ]
        \draw (-1.5,0)--(.4,0);
        \draw (1.1,0)--(1.6,0);
        \draw (-0.1,0.33) to[out=90, in=180] (.8,0);
        \draw (-0.9,-0.6)--(-0.9,0.6);
        \draw ( 1.1,-0.6)--( 1.1,0.6);
        \node[mpo](t) at (-0.9,0) {\small $Z^{a}$};   
        \node[mpo](t) at ( 1.1,0) {\small $Z^{c_n}$};
        \filldraw (.4,0) circle (.5mm);
        \node[below] at (.4,-.05) {\small $\beta$};
        \draw[fill=white, draw=black] (-.1,0) circle (.35cm) node {\small $\mathscr{C}_{ab}^{c_n}$};
    \end{tikzpicture}
    \,.
    \label{apx_eq:ZZ_string_fusion}
\end{equation}
Given these we can identify the symmetry multiplets of local order parameters of Rep($G$) symmetries. 
Commuting $\Xl{g}$ through an $R_{\Gamma}$ symmetry operators inserts a $\Gamma^g$ matrix on the virtual leg of the MPO,
\begin{equation}
    \begin{tikzpicture}[
            baseline={(0,.4)}, 
            transform shape, 
            scale=1
            ]
        \draw (-1.5, 1.0)--(1.5, 1.0);
        \draw (-1.0,-0.5)--(-1.0,1.5);
        \draw ( 0.0,-0.5)--( 0.0,1.5);
        \draw ( 1.0,-0.5)--( 1.0,1.5);
        \node[mpo](t) at (0,0) {\small $\Xl{g}$};
        \node[mpo](t) at (-1,1) {\small $Z^{\Gamma_a}$};
        \node[mpo](t) at ( 0,1) {\small $Z^{\Gamma_a}$};
        \node[mpo](t) at ( 1,1) {\small $Z^{\Gamma_a}$};
      \end{tikzpicture} 
      =
    \begin{tikzpicture}[
            baseline={(0,.4)}, 
            transform shape, 
            scale=1
            ]
        \draw (-2.0, 0.0)--(1.5, 0.0);
        \draw (-1.5,-0.5)--(-1.5,1.5);
        \draw ( 0.0,-0.5)--( 0.0,1.5);
        \draw ( 1.0,-0.5)--( 1.0,1.5);
        \node[mpo](t) at ( 0.0,1)    {\small $\Xl{g}$};
        \node[mpo](t) at (-1.5,0)   {\small $Z^{\Gamma_a}$};
        \node[mpo](t) at ( 0.0,0)   {\small $Z^{\Gamma_a}$};
        \node[mpo](t) at ( 1.0,0)   {\small $Z^{\Gamma_a}$};
        \draw[fill=white, draw=black] (-.75,0) circle (.3cm) node {\small $\Gamma_a^g$};
      \end{tikzpicture} 
      \,.
      \label{apx_eq:XR_RX}
\end{equation}
The virtual leg insertion matrix $\Gamma_a^g$ can be written as a linear combinations of the Clebsch-Gordan coefficient matrices $\mathscr{C}_{a b}^{a}$, which form a basis for linear maps $\cramped{V_{\Gamma_a} \to V_{\Gamma_a}}$,
\begin{equation}
    [\Gamma_a^g]_{\alpha \alpha'} 
    = 
    \sum_{b} 
    \sum_{\beta=1}^{d_{\Gamma_b}}
    \,\,[v_{ab}^g]_\beta [\mathscr{C}_{ab}^{a}]_{\alpha \beta}^{\alpha'}
    \,,
    \label{apx_eq:Gamma_g_expansion_fusion}
\end{equation}
We can identify the right hand side of this equation as $\Xl{g}$ at the end of half-infinite $Z^{\Gamma_b}$ string operator whose open end is contracted with the coefficients $[v_{ab}^g]_{\beta}$.

\renewcommand{\theequation}{F.\arabic{equation}} 
\setcounter{equation}{0}

\textit{Appendix F: Clebsch-Gordan Coefficients for $D_3$}---Here we illustrate the construction for the group $D_3$, using the following basis for the $E$ irrep,
\begin{equation}
    E^r = \begin{pmatrix}
        \omega & 0 \\ 0 & \omega^*
    \end{pmatrix} \equiv \Re(\omega) \mathbbm{1} +\, i \, \Im(\omega) \,\sigma^z,
    \,\,
    E^s = \begin{pmatrix}
        0 & 1 \\ 1 & 0
    \end{pmatrix} \equiv \sigma^x,
    \label{apx_eq:Er_Es}
\end{equation}
and the dual $\bar{E}$ matrices are the complex-conjugates of these.
In this basis, the matrix of Clebsch-Gordan coefficients encoding the decomposition $\cramped{E\otimes \bar{E} = A_1 \oplus A_2 \oplus E}$ is given by
\begin{equation}
    \underbrace{
    \begin{pmatrix}
        \ket{A_1,1} \\
        \ket{A_2,1} \\
        \ket{E_{\phantom{1}},1} \\
        \ket{E_{\phantom{1}},2}
    \end{pmatrix}
    }_{
        \ket{\Gamma_{c_n},\gamma}
    }
    =
    \underbrace{
    \begin{pmatrix}
        \tfrac{1}{\sqrt{2}} & 0 & 0 &  \tfrac{1}{\sqrt{2}} \\[1mm]
        \tfrac{1}{\sqrt{2}} & 0 & 0 & -\tfrac{1}{\sqrt{2}} \\[1mm]
        0 & 1 & 0 & 0 \\
        0 & 0 & 1 & 0
    \end{pmatrix}
    }_{
    [\mathscr{C}_{E\bar{E}}^{c_n}]_{\alpha\beta}^\gamma
    }
    \,
    \underbrace{
    \begin{pmatrix}
        \ket{E,1}\otimes \ket{\bar{E},1}\\
        \ket{E,1}\otimes \ket{\bar{E},2}\\
        \ket{E,2}\otimes \ket{\bar{E},1}\\
        \ket{E,2}\otimes \ket{\bar{E},2}
    \end{pmatrix}
    }_{
        \ket{E,\alpha}\otimes \ket{\bar{E},\beta}
    }
    \,,
\end{equation}
which are given by
\begin{align}
    [\mathscr{C}_{E\bar{E}}^{A_1}]_{\alpha\beta}^1 
    = 
    \tfrac{1}{\sqrt{2}}
    \scalebox{0.8}{
    $\begin{pmatrix*}
        1 & 0 \\ 0 & 1
    \end{pmatrix*}$
    }
    = 
    \frac{\delta_{\alpha\beta}}{\sqrt{2}}
    ,
    &&
    [\mathscr{C}_{E\bar{E}}^E]_{\alpha\beta}^1
    = 
    \scalebox{0.8}{
    $\begin{pmatrix*}
        0 & 1 \\ 0 & 0
    \end{pmatrix*} $
    }
    =
    \sigma^+_{\alpha\beta}
    ,
    \nonumber
    \\
    [\mathscr{C}_{E\bar{E}}^{A_2}]_{\alpha\beta}^1
    = 
    \tfrac{1}{\sqrt{2}}
    \scalebox{0.8}{
    $\begin{pmatrix}
        1 & 0 \\ 0 & -1
    \end{pmatrix}$
    }
    = \frac{\sigma^z_{\alpha\beta}}{\sqrt{2}}
    ,
    &&
     [\mathscr{C}_{E\bar{E}}^E]_{\alpha\beta}^2 
     = 
    \scalebox{0.8}{
    $\begin{pmatrix}
        0 & 0 \\ 1 & 0
    \end{pmatrix}$ 
    }
    =
    \sigma^-_{\alpha\beta}
    ,
\end{align}
These form a basis for the domain wall defects, \cref{eq:CG_MPS_defect}, which can appear in an $\ket{\bm{E}}$ ground state. Note that the trivial irrep $A_1$ corresponds to no domain wall, correspondingly it simply gives the normalization matrix $\Lambda$ from \cref{eq:Gamma_MPS_closed}.

To obtain the non-invertible symmetry multiplets such as  \cref{eq:E_r_multiplet}, the relevant matrices are
\begin{align}
    [\mathscr{C}_{E A_1}^{E}]_{\alpha 1}^{\gamma} 
    = 
    \scalebox{0.8}{
    $\begin{pmatrix*}
        1 & 0 \\ 0 & 1
    \end{pmatrix*}$
    }
    =
    \delta_{\alpha\gamma}
    ,
    \quad
    [\mathscr{C}_{E A_2}^{E}]_{\alpha 1}^{\gamma} 
    = 
    \scalebox{0.8}{
    $\begin{pmatrix*}
        1 & 0 \\ 0 & -1
    \end{pmatrix*}$
    }
    =
    \sigma^z_{\alpha\gamma}
    ,
    \nonumber
    \\
    [\mathscr{C}_{EE}^E]_{\alpha 1}^\gamma 
     = 
    \scalebox{0.8}{
    $\begin{pmatrix}
        0 & 1 \\ 0 & 0
    \end{pmatrix}$ 
    }
    =
    \sigma^+_{\alpha\gamma}
    ,
    \quad\quad
    [\mathscr{C}_{EE}^E]_{\alpha 2}^\gamma
     = 
    \scalebox{0.8}{
    $\begin{pmatrix}
        0 & 0 \\ 1 & 0
    \end{pmatrix}$ 
    }
    =
    \sigma^-_{\alpha\gamma}
    \,
    ,
    \label{apx_eq:E_to_E_fusion}
\end{align}
where $\alpha$ labels the rows and $\gamma$ labels the columns. These  matrices form a basis for maps $V_E \to V_E$, and so can be used to expand the $E^g$ matrices, \cref{apx_eq:Gamma_g_expansion_fusion}. The simple case is $E^r$, which using \cref{apx_eq:Er_Es} can be separated into an identity component and a $\sigma^z$ component, the first of which is inserted by fusion with an $A_1$ line and the second of which is inserted by fusion with an $A_2$ line, thus obtaining \cref{eq:E_r_multiplet}. The $E^s = \sigma^x$ case is more complex. Using \cref{apx_eq:Gamma_g_expansion_fusion} we see that fusing an open $E$ string with endpoint contracted with $\ket{+}=\ket{1}+\ket{2}$ into the $R_E$ MPO inserts a $\sigma^x$ defect, but it also produces an $A_1$ and $A_2$ fusion channel. This implies that three operators are mixed in this multiplet---a local $\Xl{s}$ operator, an $\Xl{s}$ operator at the end of an $E$ string with endpoint $\ket{+}$, and an $\Xl{s}$ operator at a junction between an $E$ string with endpoint $\ket{-}$ and an $A_2$ string. More detailed derivations can be found in~\cite{SM}.


\clearpage

\onecolumngrid

\pagenumbering{roman}

\renewcommand{\theequation}{S.\arabic{equation}} 
\setcounter{equation}{0}

\begin{center}
    \textbf{\large Supplementary Material}
\end{center}

\section{\textit{G}-Qudits}

A single $G$-qudit Hilbert space is spanned by the group elements of a finite group $G$
\begin{equation}
    \mathscr{H} = \mathbb{C}[G] = \text{span}_{\mathbb{C}}\{\ket{g} \vert\, g \in G\},
    \label{SM_eq:Hilbert_space}
\end{equation}
Its dimension is equal to the number of group elements $\vert G \vert$. 
We have the following generalized Pauli operators acting on this space,\footnote{
    The $\smash{\Xl{g}}$ operators furnish the ``left regular representation'' of $G$, while $\smash{\Xr{g}}$ furnish the ``right regular representation''. According to Cayley's theorem, every finite group $G$ is a subgroup of $\cramped{S_{\vert G\vert}}$, the permutation group of $\vert G \vert$ elements. The regular representations are precisely the restrictions of the fundamental representation of $\cramped{S_{\vert G \vert}}$ to the $G$ subgroup. 
    In the $\ket{g}$ basis, the $X$ operators are precisely these permutation matrices. 
}
\begin{equation}
    \Xl{g} \ket{h} = \ket{gh}, 
    \quad  
    \Xr{g} \ket{h} = \ket{hg^{-1}},
    \quad 
    Z^{\Gamma}_{\alpha\beta}\ket{g} = \Gamma^{g}_{\alpha\beta} \ket{g}, \qquad g,h \in G,
    \label{SM_eq:Xl_Xr_Z}
\end{equation}
where $\Gamma$ is a $d_{\Gamma}$-dimensional irreducible representation of $G$, and the indices $\alpha,\beta = 1\ldots d_{\Gamma}$ index the unitary matrix elements of the $\Gamma^g$ in a particular basis.
These operators can also be written as
\begin{equation}
    \Xl{g} = \sum_h \ket{gh}\bra{h} ,\qquad \Xr{g} = \sum_h \ket{h g^{-1}}\bra{h}, \qquad Z^{\Gamma}_{\alpha\beta} = \sum_{g} \Gamma_{\alpha\beta}^{g} \ket{g}\bra{g}.
\end{equation}
The Hermitian conjugates are given by,
\begin{equation}
    {Z^{\Gamma}_{\alpha\beta}}^\dagger = Z^{\Gamma^*}_{\alpha\beta}, 
    \qquad 
    {\Xl{g}}^\dagger = \Xl{g^{-1}},
    \qquad 
    {\Xr{g}}^\dagger = \Xr{g^{-1}}.
\end{equation}
where $\Gamma^*$ denotes the complex-conjugate representation.
The $\Xl{g}$ operators do not commute with each other if the group is non-Abelian, rather they commute up to conjugation,
\begin{equation}
    [\Xl{g},\Xl{h}] = \Xl{gh} - \Xl{hg}, \qquad \Xl{g}\Xl{h} = \Xl{h}\Xl{h^{-1}gh}.
\end{equation}
Left and right multiplication commute, 
\begin{equation}
    [\Xl{g},\Xr{h}] = 0.
\end{equation}
The algebra of $X$ and $Z$ operators is 
\begin{equation}
    Z_{\alpha\beta}^{\Gamma} \Xl{g}  =  \Gamma_{\alpha\gamma}^g \,\Xl{g}Z^{\Gamma}_{\gamma\beta},
    \qquad
    Z_{\alpha\beta}^{\Gamma} \Xr{g}  =   \,\Xr{g}Z^{\Gamma}_{\alpha\gamma}\Gamma_{\gamma\beta}^{g^{-1}}.
    \label{SM_eq:ZX_G}
\end{equation}

\subsection{Dual Basis}

Given an irrep $\Gamma$, we introduce
\begin{equation}
    \ket{\Gamma_{\alpha\beta}} = \sqrt{\frac{d_{\Gamma}}{\vert G \vert}} \sum_{g \in G} \Gamma_{\alpha\beta}^g \ket{g}.
    \label{SM_eq:ket_Gamma_ab}
\end{equation}
Doing that for all ireps (to be labeled below by latin indices),
one forms a dual orthonormal basis of the G-qudit Hilbert space,
\begin{equation}
    \braket{\Gamma_{a,\alpha\beta}\vert\Gamma_{b,\alpha'\beta'}}
    =
    \frac{\sqrt{d_{\Gamma_a}d_{\Gamma_b}}}{\vert G\vert}
    \sum_g 
    \Gamma^{g*}_{a,\alpha\beta}
    \Gamma^{g}_{b,\alpha'\beta'} 
    = \delta_{ab}\delta_{\alpha\alpha'}\delta_{\beta\beta'}  ,
    \label{SM_eq:GOT}
\end{equation}
which is precisely the statement of the Great Orthogonality Theorem of finite groups.
For a finite group $G$, we have the following identity,
\begin{equation}
    \sum_{\Gamma} d_{\Gamma}^2 = \vert G \vert,
\end{equation}
which ensures that the number of the dual basis states $\ket{\Gamma_{a,\alpha\beta}}$ is equal to the number of the basis states $\ket{g}$. Clearly, the two bases are related by a unitary transformation.
While the $Z$ operators are diagonal in the $\ket{g}$ basis, the $X$ operators are block-diagonal in the dual basis,
\begin{equation}
    \Xl{g} = \sum_{\Gamma} \sum_{\alpha,\beta=1}^{d_{\Gamma}} \Gamma_{\alpha\alpha'}^{g^{-1}} \ket{\Gamma_{\alpha'\beta}}\bra{\Gamma_{\alpha\beta}},
    \qquad
    \Xr{g} = \sum_{\Gamma} \sum_{\alpha,\beta=1}^{d_{\Gamma}} \Gamma_{\beta'\beta}^{g}\ket{\Gamma_{\alpha\beta'}}\bra{\Gamma_{\alpha\beta}}  .
    \label{SM_eq:X_action_Gamma}
\end{equation}
That is, $\Xl{g}$ acts only on the left index while $\Xr{g}$ acts on the right index. For Abelian groups, $d_{\Gamma} = 1$ for every irrep, so there are no indices and the $X$ operators are diagonal in this basis. 
For non-Abelian groups, the $X$ operators do not commute with one another, so they can only be block-diagonalized.

\subsection{MPS and MPO Representation}

It is very useful to express the $Z^{\Gamma}_{\alpha\beta}$ operators (\cref{SM_eq:Xl_Xr_Z}) as matrix product operators (MPOs) and the $\ket{\Gamma_{\alpha\beta}}$ states (\cref{SM_eq:ket_Gamma_ab}) as matrix product state (MPS) tensors, 
\begin{equation}
    Z_{\alpha\beta}^\Gamma
    \equiv
    \begin{tikzpicture}[
        baseline={(0,0)}, 
        transform shape, 
        scale=0.9
        ]
        \draw (-0.5,0)--(0.5,0);
        \draw (0,-0.5)--(0,0.5);
        \node[mpo](t) at (0,0) {\small $Z^\Gamma$};
        \filldraw (-.5,0) circle (.5mm);
        \filldraw (+.5,0) circle (.5mm);
        \node[left]  at (-0.5, 0  ) {\small $\alpha$};
        \node[right] at ( 0.5, 0  ) {\small $\beta$};
        \node[above] at ( 0  , 0.5) {\small $\ket{g}$};
        \node[below] at ( 0  ,-0.5) {\small $\bra{g}$};
    \end{tikzpicture} 
    \qquad
    \text{and}
    \qquad
    \ket{\Gamma_{\alpha\beta}}
    \equiv
    \begin{tikzpicture}[baseline={(0,0.1)}]
        \draw (-0.5, 0.0)--(0.5,0.0);
        \draw ( 0.0, 0.0)--(0.0,0.5);
        \node[mpo](t) at (0,0) {\small $\ket{\Gamma}$};
        \filldraw (-0.5,0) circle (.5mm);
        \filldraw (+0.5,0) circle (.5mm);
        \node[left]  at (-.5,0) {\small $\alpha$};
        \node[right] at (+.5,0) {\small $\beta$};
        \node[above] at (0,0.5) {\small $\ket{g}$};
    \end{tikzpicture} 
    .
\end{equation}
In this form, the algebra of the $X$ and $Z$ operators, \cref{SM_eq:ZX_G} can be expressed as
\begin{equation}
    \begin{tikzpicture}[baseline={(0,.45)}]
        \draw (-0.5,1)--(0.5,1);
        \draw (0,-0.5)--(0,1.5);
        \node[mpo](t) at (0,0) {\small $\Xl{g}$};
        \node[mpo](t) at (0,1) {\small $Z^\Gamma$};
        \filldraw (-.5,1) circle (.5mm);
        \filldraw (+.5,1) circle (.5mm);
        \node[left]  at (-.5,1) {\small $\alpha$};
        \node[right] at (+.5,1) {\small $\beta$};
    \end{tikzpicture} 
    =
    \begin{tikzpicture}[baseline={(0,.45)}]
        \draw (-1.3,0)--(0.5,0);
        \draw (0,-0.5)--(0,1.5);
        \node[mpo](t) at (0,1) {\small $\Xl{g}$};
        \node[mpo](t) at (0,0) {\small $Z^\Gamma$};
        \draw[fill=white, draw=black] (-0.8,0) circle (.3cm) node {$\Gamma^g$};
        \filldraw (-1.3,0) circle (.5mm);
        \filldraw (+0.5,0) circle (.5mm);
        \node[left]  at (-1.3,0) {\small $\alpha$};
        \node[right] at (+0.5,0) {\small $\beta$};
    \end{tikzpicture} 
    \quad
    ,
    \qquad
    \qquad
    \begin{tikzpicture}[baseline={(0,.45)}]
        \draw (-0.5,1)--(0.5,1);
        \draw (0,-0.5)--(0,1.5);
        \node[mpo](t) at (0,0) {\small $\Xr{g}$};
        \node[mpo](t) at (0,1) {\small $Z^\Gamma$};
        \filldraw (-.5,1) circle (.5mm);
        \filldraw (+.5,1) circle (.5mm);
        \node[left]  at (-.5,1) {\small $\alpha$};
        \node[right] at (+.5,1) {\small $\beta$};
    \end{tikzpicture} 
    =
    \begin{tikzpicture}[baseline={(0,.45)}]
        \draw (-0.5, 0.0)--(1.3,0.0);
        \draw ( 0.0,-0.5)--(0.0,1.5);
        \node[mpo](t) at (0,1) {\small $\Xr{g}$};
        \node[mpo](t) at (0,0) {\small $Z^\Gamma$};
        \draw[fill=white, draw=black] (0.8,0) circle (.3cm) node {$\,\overline{\Gamma}^{\raisemath{-2pt}{\,\smash{g}}}$};
        \filldraw (-0.5,0) circle (.5mm);
        \filldraw (+1.3,0) circle (.5mm);
        \node[left]  at (-0.5,0) {\small $\alpha$};
        \node[right] at (+1.3,0) {\small $\beta$};
    \end{tikzpicture} 
    \quad
    ,
    \label{SM_eq:ZX_XZ}
\end{equation}
where $\overline{\Gamma}^{\raisemath{-2pt}{\,\smash{g}}} = \Gamma^{g^{-1}}$. 
In words, commuting $\Xl{g}$ past $Z^{\Gamma}_{\alpha\beta}$ multiplies the left virtual leg by the representation matrix $\Gamma^g$, and analogously for $\smash{\Xr{g}}$. 
The action of the $X$ operators on the dual basis MPS tensors, \cref{SM_eq:X_action_Gamma}, similarly multiplies the virtual legs
\begin{equation}
    \begin{tikzpicture}[baseline={(0,0.1)}]
        \draw (-0.5, 0.0)--(0.5,0.0);
        \draw ( 0.0, 0.0)--(0.0,1.5);
        \node[mpo](t) at (0,0) {\small $\ket{\Gamma}$};
        \node[mpo](t) at (0,1) {\small $\Xl{g}$};
        \filldraw (-0.5,0) circle (.5mm);
        \filldraw (+0.5,0) circle (.5mm);
        \node[left]  at (-.5,0) {\small $\alpha$};
        \node[right] at (+.5,0) {\small $\beta$};
    \end{tikzpicture} 
    \quad
    =
    \quad
    \begin{tikzpicture}[baseline={(0,0.1)}]
        \draw (-1.5, 0.0)--(0.5,0.0);
        \draw ( 0.0, 0.0)--(0.0,1.5);
        \draw[fill=white, draw=black] (-.9,0) circle (.3cm) node {\small $\,\overline{\Gamma}^{\raisemath{-2pt}{\,\smash{g}}}$};
        \node[mpo](t) at (0,0) {\small $\ket{\Gamma}$};
        \filldraw (-1.5,0) circle (.5mm);
        \filldraw (+0.5,0) circle (.5mm);
        \node[left]  at (-1.5,0) {\small $\alpha$};
        \node[right] at (+0.5,0) {\small $\beta$};
    \end{tikzpicture} 
    \,\,
    ,
    \qquad
    \qquad
    \begin{tikzpicture}[baseline={(0,0.1)}]
        \draw (-0.5, 0.0)--(0.5,0.0);
        \draw ( 0.0, 0.0)--(0.0,1.5);
        \node[mpo](t) at (0,0) {\small $\ket{\Gamma}$};
        \node[mpo](t) at (0,1) {\small $\Xr{g}$};
        \filldraw (-0.5,0) circle (.5mm);
        \filldraw (+0.5,0) circle (.5mm);
        \node[left]  at (-.5,0) {\small $\alpha$};
        \node[right] at (+.5,0) {\small $\beta$};
    \end{tikzpicture} 
    \quad
    =
    \quad
    \begin{tikzpicture}[baseline={(0,0.1)}]
        \draw (-0.5, 0.0)--(1.3,0.0);
        \draw ( 0.0, 0.0)--(0.0,1.3);
        \draw[fill=white, draw=black] (.8,0) circle (.3cm) node {\small $\Gamma^g$};
        \node[mpo](t) at (0,0) {\small $\ket{\Gamma}$};
        \filldraw (-0.5,0) circle (.5mm);
        \filldraw (+1.3,0) circle (.5mm);
        \node[left]  at (-0.5,0) {\small $\alpha$};
        \node[right] at (+1.3,0) {\small $\beta$};
    \end{tikzpicture} 
    \quad
    .
\end{equation}
As an MPS statement, the orthogonality of the dual basis MPS tensors, \cref{SM_eq:GOT}, can be written as 
\begin{equation}
    \begin{tikzpicture}[baseline={(0,0.5)}]
        \draw ( 0.0, 0.0)--(0.0,1);
        \draw (-0.5, 0.0)--(0.5,0.0);
        \node[mpo](t) at (0,0) {\small $\ket{\Gamma'}$};
        \filldraw (-0.5,0) circle (.5mm);
        \filldraw (+0.5,0) circle (.5mm);
        \node[left]  at (-.5,0) {\small $\alpha'$};
        \node[right] at (+.5,0) {\small $\beta'$};
        \draw (-0.5, 1.0)--(0.5,1.0);
        \node[mpo](t) at (0,1) {\small $\bra{\Gamma}$};
        \filldraw (-0.5,1) circle (.5mm);
        \filldraw (+0.5,1) circle (.5mm);
        \node[left]  at (-.5,1) {\small $\alpha$};
        \node[right] at (+.5,1) {\small $\beta$};
    \end{tikzpicture} 
    \quad
    =
    \quad
    \delta_{\Gamma,\Gamma'} 
    \quad
    \begin{tikzpicture}[baseline={(0,0.5)}]
        \draw[rounded corners] (-.5,0)--(-.25,0)--(-.25,1)--(-.5,1);
        \draw[rounded corners] (.5,0)--(.25,0)--(.25,1)--(.5,1);
        \filldraw (-0.5,0) circle (.5mm);
        \filldraw (+0.5,0) circle (.5mm);
        \filldraw (-0.5,1) circle (.5mm);
        \filldraw (+0.5,1) circle (.5mm);
        \node[left]  at (-.5,0) {\small $\alpha'$};
        \node[right] at (+.5,0) {\small $\beta'$};
        \node[left]  at (-.5,1) {\small $\alpha$};
        \node[right] at (+.5,1) {\small $\beta$}; 
    \end{tikzpicture}  \, .
    \label{SM_eq:orthogonality_MPS}
\end{equation}

\section{Exact SSB Ground States at \texorpdfstring{$h=0$}{h=0}}

For the Rep($G$)-symmetric transverse-field Hamiltonian discussed in the main text, in the limit $h=0$ we can construct the exact Rep($G$) symmetry-broken states.
The Hamiltonian in this limit generally takes the form $H = \sum_i \sum_g J_g \Xr{g}_i\Xl{g}_{i+1}$. Each two-site term acts in the dual basis as 
\begin{equation}
    \begin{tikzpicture}[baseline={(0,0.1)}]
        \draw (-0.5, 0.0)--(0.5,0.0);
        \draw ( 0.0, 0.0)--(0.0,1.5);
        \node[mpo](t) at (0,0) {\small $\ket{\Gamma}$};
        \node[mpo](t) at (0,1) {\small $\Xr{g}$};
        \filldraw (-0.5,0) circle (.5mm);
        \filldraw (+0.5,0) circle (.5mm);
        \node[left]  at (-.5,0) {\small $\cdots\,\alpha$};
        \node[right] at (+.5,0) {\small $\beta$};
    \end{tikzpicture} 
    \begin{tikzpicture}[baseline={(0,0.1)}]
        \draw (-0.5, 0.0)--(0.5,0.0);
        \draw ( 0.0, 0.0)--(0.0,1.5);
        \node[mpo](t) at (0,0) {\small $\ket{\Gamma}$};
        \node[mpo](t) at (0,1) {\small $\Xl{g}$};
        \filldraw (-0.5,0) circle (.5mm);
        \filldraw (+0.5,0) circle (.5mm);
        \node[left]  at (-.5,0) {\small $\gamma$};
        \node[right] at (+.5,0) {\small $\delta\,\cdots$};
    \end{tikzpicture} 
    =
    \begin{tikzpicture}[baseline={(0,0.1)}]
        \draw (-0.5, 0.0)--(1.3,0.0);
        \draw ( 0.0, 0.0)--(0.0,1.3);
        \draw[fill=white, draw=black] (.8,0) circle (.3cm) node {\small $\Gamma^g$};
        \node[mpo](t) at (0,0) {\small $\ket{\Gamma}$};
        \filldraw (-0.5,0) circle (.5mm);
        \filldraw (+1.3,0) circle (.5mm);
        \node[left]  at (-0.5,0) {\small $\cdots\,\alpha$};
        \node[right] at (+1.3,0) {\small $\beta$};
    \end{tikzpicture} 
    \begin{tikzpicture}[baseline={(0,0.1)}]
        \draw (-1.5, 0.0)--(0.5,0.0);
        \draw ( 0.0, 0.0)--(0.0,1.5);
        \draw[fill=white, draw=black] (-.9,0) circle (.3cm) node {\small $\,\overline{\Gamma}^{\raisemath{-2pt}{\,\smash{g}}}$};
        \node[mpo](t) at (0,0) {\small $\ket{\Gamma}$};
        \filldraw (-1.5,0) circle (.5mm);
        \filldraw (+0.5,0) circle (.5mm);
        \node[left]  at (-1.5,0) {\small $\gamma$};
        \node[right] at (+0.5,0) {\small $\delta\,\cdots$};
    \end{tikzpicture} 
\end{equation}
From this it becomes clear that states with neighboring indices contracted are eigenstates with eigenvalue 1,
\begin{equation}
    \begin{tikzpicture}[baseline={(0,0.1)}]
        \draw (-0.5, 0.0)--(1.5,0.0);
        \draw ( 0.0, 0.0)--(0.0,1.5);
        \node[mpo](t) at (0,0) {\small $\ket{\Gamma}$};
        \node[mpo](t) at (0,1) {\small $\Xr{g}$};
        \filldraw (-0.5,0) circle (.5mm);
        \node[left]  at (-.5,0) {\small $\cdots\,\alpha$};
        \draw ( 1.0, 0.0)--(1.0,1.5);
        \node[mpo](t) at (1,0) {\small $\ket{\Gamma}$};
        \node[mpo](t) at (1,1) {\small $\Xl{g}$};
        \filldraw (+1.5,0) circle (.5mm);
        \node[right] at (+1.5,0) {\small $\beta\,\cdots$};
    \end{tikzpicture} 
    =
    \begin{tikzpicture}[baseline={(0,0.1)}]
        \draw (-0.5, 0.0)--(1.5,0.0);
        \draw ( 0.0, 0.0)--(0.0,1.5);
        \node[mpo](t) at (0,0) {\small $\ket{\Gamma}$};
        \filldraw (-0.5,0) circle (.5mm);
        \node[left]  at (-.5,0) {\small $\cdots\,\alpha$};
        \draw ( 1.0, 0.0)--(1.0,1.5);
        \node[mpo](t) at (1,0) {\small $\ket{\Gamma}$};
        \filldraw (+1.5,0) circle (.5mm);
        \node[right] at (+1.5,0) {\small $\beta\,\cdots$};
    \end{tikzpicture} 
\end{equation}
It follows then that if all $J_g > 0$, then the normalized ground states are given by
\begin{equation}
    \ket{\bm{\Gamma}} = 
    \begin{tikzpicture}[baseline={(0,-0.1)}]
        \draw (-.5,0)--(2.5,0);
        \draw (0,0)--(0,0.7);
        \draw (1,0)--(1,0.7);
        \draw (2,0)--(2,0.7);
        \node[mpo](t) at (0,0) {\small $\ket{\Gamma}$};
        \node[mpo](t) at (1,0) {\small $\ket{\Gamma}$};
        \node[mpo](t) at (2,0) {\small $\ket{\Gamma}$};
        \filldraw[fill=white] (+0.5,0) circle (.7mm);
        \filldraw[fill=white] (+1.5,0) circle (.7mm);
        \node[left]  at (-0.5,0) {$\dots$};
        \node[right] at (+2.5,0) {$\dots$};
    \end{tikzpicture}  
    \equiv 
    \sum_{\{\alpha_i\}}
    \bigotimes_i \frac{1}{\sqrt{d_{\Gamma}}}\ket{\Gamma_{\alpha_{i}\alpha_{i+1}}}_i \, .
    \label{SM_eq:Gamma_GS}
\end{equation}
Here we have included one normalization factor for each contraction of neighboring virtual indices, which are indicated as insertions on each virtual bond of the MPS by a white circle representing the diagonal matrix
\begin{equation}
    \Lambda_{\alpha\beta}
    \equiv
    \begin{tikzpicture}[baseline={(0,-0.1)}]
        \draw (-.5,0)--(.5,0);
        \filldraw[fill=white] (0,0) circle (.7mm);
        \filldraw (-0.5,0) circle (.5mm);
        \filldraw ( 0.5,0) circle (.5mm);
        \node[left]  at (-.5,0) {\small $\alpha$};
        \node[right] at ( .5,0) {\small $\beta$};
    \end{tikzpicture}
    =
    \frac{1}{\sqrt{d_{\Gamma}}}\,\delta_{\alpha\beta}
    \,.
\end{equation}
The MPS \cref{SM_eq:Gamma_GS} is in canonical form, where the matrix $\Lambda$ contains the $d_\Gamma$ degenerate Schmidt eigenvalues. 
The normalization of the ground state can be checked
\begin{equation}
   \langle \bm \Gamma \ket{\bm \Gamma} = 
    \begin{tikzpicture}[baseline={(0,0.4)}]
        \draw (-.5,0)--(2.5,0);
        \draw (-.5,1)--(2.5,1);
        \draw (0,0)--(0,1);
        \draw (1,0)--(1,1);
        \draw (2,0)--(2,1);
        \node[mpo](t) at (0,0) {\small $\ket{\Gamma}$};
        \node[mpo](t) at (1,0) {\small $\ket{\Gamma}$};
        \node[mpo](t) at (2,0) {\small $\ket{\Gamma}$};
        \node[mpo](t) at (0,1) {\small $\bra{\Gamma}$};
        \node[mpo](t) at (1,1) {\small $\bra{\Gamma}$};
        \node[mpo](t) at (2,1) {\small $\bra{\Gamma}$};
        \filldraw[fill=white] (0.5,0) circle (.7mm);
        \filldraw[fill=white] (1.5,0) circle (.7mm);
        \filldraw[fill=white] (0.5,1) circle (.7mm);
        \filldraw[fill=white] (1.5,1) circle (.7mm);
        \node[left]  at (-0.5,0) {$\dots$};
        \node[right] at (+2.5,0) {$\dots$};
        \node[left]  at (-0.5,1) {$\dots$};
        \node[right] at (+2.5,1) {$\dots$};
    \end{tikzpicture}  
    \equiv 
    \begin{tikzpicture}[baseline={(0,0.4)}]
        \node[left] at (-1,0) {$\dots$};
        \node[left] at (-1,1) {$\dots$};
        \draw[rounded corners] (-1,0)--(-.75,0)--(-.75,1)--(-1,1);
        \draw[rounded corners] (0,0)--(0.25,0)--(0.25,1)--(-0.25,1)--(-0.25,0)--(0,0);
        \draw[rounded corners] (1,0)--(1.25,0)--(1.25,1)--( 0.75,1)--( 0.75,0)--(1,0);
        \draw[rounded corners] (2,0)--(1.75,0)--(1.75,1)--(2,1);
        \filldraw[fill=white] (0,0) circle (.7mm);
        \filldraw[fill=white] (1,0) circle (.7mm);
        \filldraw[fill=white] (0,1) circle (.7mm);
        \filldraw[fill=white] (1,1) circle (.7mm);
        \node[right] at (2,0) {$\dots$};
        \node[right] at (2,1) {$\dots$};
    \end{tikzpicture}  
    =
    1 \, ,
\end{equation}
where the closed loops are traces of $\Lambda^2$,
\begin{equation}
    \begin{tikzpicture}[baseline={(0,0.4)}]
        \draw[rounded corners] (0,0)--(0.25,0)--(0.25,1)--(-0.25,1)--(-0.25,0)--(0,0);
        \filldraw[fill=white] (0,0) circle (.7mm);
        \filldraw[fill=white] (0,1) circle (.7mm);
    \end{tikzpicture}  
    =
    \Tr[\Lambda^2]
    =
    \frac{1}{d_{\Gamma}} \sum_{\alpha=1}^{d_{\Gamma}} \delta_{\alpha\alpha}
    =
    1\, .
\end{equation}
We can easily compute the expectations of local operators in these states because they factorize exactly, in particular for the local order parameters we have
\begin{equation}
    \bra{\bm \Gamma} \Xl{g}_i \ket{\bm \Gamma} 
    = 
    \begin{tikzpicture}[baseline={(0,1)}]
        \draw (-.5,0)--(2.5,0);
        \draw (-.5,2)--(2.5,2);
        \draw (0,0)--(0,2);
        \draw (1,0)--(1,2);
        \draw (2,0)--(2,2);
        \node[mpo](t) at (0,0)   {\small $\ket{\Gamma}$};
        \node[mpo](t) at (1,0)   {\small $\ket{\Gamma}$};
        \node[mpo](t) at (2,0)   {\small $\ket{\Gamma}$};
        \node[mpo](t) at (1,1)   {\small $\Xl{g}$};
        \node[mpo](t) at (0,2)   {\small $\bra{\Gamma}$};
        \node[mpo](t) at (1,2)   {\small $\bra{\Gamma}$};
        \node[mpo](t) at (2,2)   {\small $\bra{\Gamma}$};
        \node[left]  at (-0.4,0) {\small $\dots$};
        \node[right] at (+2.4,0) {\small $\dots$};
        \node[left]  at (-0.4,2) {\small $\dots$};
        \node[right] at (+2.4,2) {\small $\dots$};
        \filldraw[fill=white] (0.5,0) circle (.7mm);
        \filldraw[fill=white] (1.5,0) circle (.7mm);
        \filldraw[fill=white] (0.5,2) circle (.7mm);
        \filldraw[fill=white] (1.5,2) circle (.7mm);
    \end{tikzpicture}  
    = 
    \begin{tikzpicture}[baseline={(0,1)}]
        \draw (-.5,0)--(3.5,0);
        \draw (-.5,2)--(3.5,2);
        \draw (0,0)--(0,2);
        \draw (2,0)--(2,2);
        \draw (3,0)--(3,2);
        \node[mpo](t) at (0.0,0) {\small $\ket{\Gamma}$};
        \node[mpo](t) at (2,0)   {\small $\ket{\Gamma}$};
        \node[mpo](t) at (3,0)   {\small $\ket{\Gamma}$};
        \node[mpo](t) at (0.0,2) {\small $\bra{\Gamma}$};
        \node[mpo](t) at (2,2)   {\small $\bra{\Gamma}$};
        \node[mpo](t) at (3,2)   {\small $\bra{\Gamma}$};
        \node[left]  at (-0.4,0) {\small $\dots$};
        \node[right] at (+3.44,0) {\small $\dots$};
        \node[left]  at (-0.4,2) {\small $\dots$};
        \node[right] at (+3.44,2) {\small $\dots$};
        \filldraw[fill=white] (0.6,0) circle (.7mm);
        \filldraw[fill=white] (2.5,0) circle (.7mm);
        \filldraw[fill=white] (1.0,2) circle (.7mm);
        \filldraw[fill=white] (2.5,2) circle (.7mm);
        \draw[fill=white, draw=black] (1.2,0) circle (.3cm) node {\small $\,\overline{\Gamma}^{\raisemath{-2pt}{\,\smash{g}}}$};
    \end{tikzpicture} 
    =
    \begin{tikzpicture}[baseline={(0,.4)}]
        \draw[rounded corners] (-1,0)--(-.75,0)--(-.75,1)--(-1,1);
        \draw[rounded corners] (0,0)--(0.5,0)--(0.5,1)--(-0.5,1)--(-0.5,0)--(0,0);
        \draw[rounded corners] (1,0)--(1.25,0)--(1.25,1)--( 0.75,1)--( 0.75,0)--(1,0);
        \draw[rounded corners] (1.75,0)--(1.5,0)--(1.5,1)--(1.75,1);
        \filldraw[fill=white] (-0.5,0.5) circle (.7mm);
        \filldraw[fill=white] (1,0) circle (.7mm);
        \filldraw[fill=white] (0,1) circle (.7mm);
        \filldraw[fill=white] (1,1) circle (.7mm);
        \node[left]  at (-0.9,0) {\small $\dots$};
        \node[left]  at (-0.9,1) {\small $\dots$};
        \node[right] at (1.65,0) {\small $\dots$};
        \node[right] at (1.65,1) {\small $\dots$};
        \draw[fill=white, draw=black] (0,0) circle (.3cm) node {\small $\,\overline{\Gamma}^{\raisemath{-2pt}{\,\smash{g}}}$};
    \end{tikzpicture}  
    =
    \frac{\Tr[\Gamma^{g^{-1}}]}{d_{\Gamma}}.
\end{equation}
Similarly, $\bra{\bm{\Gamma}}\Xr{g}_i \ket{\bm{\Gamma}} = \Tr[\Gamma^g]/d_\Gamma$.

\section{Derivation of Non-Invertible Symmetry Multiplets of \texorpdfstring{$D_3$}{D3}}

When pulling an $\Xl{g}$ operator through the non-invertible $R_\Gamma$ MPO, using \cref{SM_eq:ZX_XZ}, a $\Gamma^g$ matrix is inserted on the virtual bond,
\begin{equation}
    \begin{tikzpicture}[baseline={(0,.45)}]
        \draw (-1.5, 1.0)--(1.5, 1.0);
        \draw (-1.0,-0.5)--(-1.0,1.5);
        \draw ( 0.0,-0.5)--( 0.0,1.5);
        \draw ( 1.0,-0.5)--( 1.0,1.5);
        \node[mpo](t) at (0,0) {\small $\Xl{g}$};
        \node[mpo](t) at (-1,1) {\small $Z^\Gamma$};
        \node[mpo](t) at ( 0,1) {\small $Z^\Gamma$};
        \node[mpo](t) at ( 1,1) {\small $Z^\Gamma$};
        \node[left]  at (-1.5,1) {$\dots$};
        \node[right] at ( 1.5,1) {$\dots$};
      \end{tikzpicture} 
      =
    \begin{tikzpicture}[baseline={(0,.45)}]
        \draw (-2.5, 0.0)--(1.5, 0.0);
        \draw (-2.0,-0.5)--(-2.0,1.5);
        \draw ( 0.0,-0.5)--( 0.0,1.5);
        \draw ( 1.0,-0.5)--( 1.0,1.5);
        \node[mpo](t) at (0,1)    {\small $\Xl{g}$};
        \node[mpo](t) at (-2,0)   {\small $Z^\Gamma$};
        \node[mpo](t) at ( 0,0)   {\small $Z^\Gamma$};
        \node[mpo](t) at ( 1,0)   {\small $Z^\Gamma$};
        \draw[fill=white, draw=black] (-1,0) circle (.3cm) node {\small $\Gamma^g$};        
        \node[left]   at (-2.5,0) {$\dots$};
        \node[right]  at ( 1.5,0) {$\dots$};
      \end{tikzpicture} .
      \label{SM_eq:XGamma_GammaX}
\end{equation}
If $d_{\Gamma}=1$ then $R_\Gamma$ is an invertible operator and $\Gamma^g$ is just a phase.
If $d_{\Gamma}>1$ then $R_{\Gamma}$ is non-invertible and the $\Gamma^g$ insertion is non-trivial. 
This insertion can be decomposed using the Clebsch-Gordan coefficients,
\begin{equation}
    [\Gamma_a^g]_{\alpha \alpha'} 
    = 
    \sum_{b} 
    \sum_{\beta=1}^{d_{\Gamma_b}}
    \,\,[\Gamma_a^g]_{(b,\beta)}[\mathscr{C}_{ab}^{a}]_{\alpha \beta}^{\alpha'}
    \,,
    \label{SM_eq:Gamma_g_expansion_fusion}
\end{equation}
to relate it to a half-infinite symmetry string operator, using
\begin{equation}
    \begin{tikzpicture}[
        baseline={(0,0.5)}, 
        transform shape, 
        scale=1
        ]
        \draw (0.4,0.0)--(2.6,0.0);
        \draw (1.4,0.8)--(2.6,0.8);
        \draw ( 1.0,-0.5)--( 1.0,1.3);
        \draw ( 2.0,-0.5)--( 2.0,1.3);
        \node[mpo](t) at (2.0,0.8) {\small $Z^{b}$};
        \node[mpo](t) at (1.0,0.0) {\small $Z^{a}$};
        \node[mpo](t) at (2.0,0.0) {\small $Z^{a}$};
        \filldraw (1.4,0.8) circle (.5mm);
        \node[above] at (1.4,.87) {\small $\beta$};
        \node[left]  at (0.4,0.0) {\small $\dots$};
        \node[right] at (2.6,0.0) {\small $\dots$};
        \node[right] at (2.6,0.8) {\small $\dots$};
    \end{tikzpicture}
    =
    \sum_{c \in a\otimes b}
    \sum_{n=1}^{N_{ab}^c} 
    \,
    \begin{tikzpicture}[
        baseline={(0,-0.1)}, 
        transform shape, 
        scale=1
        ]
        \draw (-1.5,0)--(.4,0);
        \draw (1.1,0)--(1.6,0);
        \draw (-0.1,0.33) to[out=90, in=180] (.8,0);
        \draw (-0.9,-0.6)--(-0.9,0.6);
        \draw ( 1.1,-0.6)--( 1.1,0.6);
        \node[mpo](t) at (-0.9,0) {\small $Z^{a}$};   
        \node[mpo](t) at ( 1.1,0) {\small $Z^{c_n}$};
        \filldraw (.4,0) circle (.5mm);
        \node[below] at (.4,-.05) {\small $\beta$};
        \draw[fill=white, draw=black] (-.1,0) circle (.35cm) node {\small $\mathscr{C}_{ab}^{c_n}$};
        \node[left]  at (-1.5,0) {\small $\dots$};
        \node[right] at ( 1.6,0) {\small $\dots$};
    \end{tikzpicture}
    \,.
    \label{SM_eq:ZZ_string_fusion}
\end{equation}
Here we illustrate the mechanism with the dihedral group $D_3$ generated by $r,s$ with $r^3 = s^2 = e$ and $srs=r^{-1}$. 
The non-trivial $E$ representation generated by the matrices
\begin{equation}
    E^r = \begin{pmatrix}
        \omega & 0 \\ 0 & \omega^*
    \end{pmatrix} \equiv \Re(\omega) \mathbbm{1} +\, i \, \Im(\omega) \,\sigma^z,
    \qquad
    E^s = \begin{pmatrix}
        0 & 1 \\ 1 & 0
    \end{pmatrix} \equiv \sigma^x,
    \label{SM_eq:Er_Es}
\end{equation}
where $\omega = \exp(i2\pi /3)$.
In this basis we have the following relevant Clebsch-Gordan coefficients
\begin{equation}
    [\mathscr{C}_{E A_1}^{E}]_{\alpha 1}^{\gamma} 
    = 
    \begin{pmatrix*}
        1 & 0 \\ 0 & 1
    \end{pmatrix*}
    =
    \mathbbm{1}
    ,
    \qquad
    [\mathscr{C}_{E A_2}^{E}]_{\alpha 1}^{\gamma} 
    = 
    \begin{pmatrix*}
        1 & 0 \\ 0 & -1
    \end{pmatrix*}
    =
    \sigma^z
    ,
    \qquad
    [\mathscr{C}_{EE}^E]_{\alpha 1}^\gamma 
    = 
    \begin{pmatrix}
        0 & 1 \\ 0 & 0
    \end{pmatrix}
    =
    \sigma^+
    ,
    \qquad
    [\mathscr{C}_{EE}^E]_{\alpha 2}^\gamma
    = 
    \begin{pmatrix}
        0 & 0 \\ 1 & 0
    \end{pmatrix}
    =
    \sigma^-
    \,.
    \label{SM_eq:E_to_E_fusion}
\end{equation}
where $\alpha$ indexes rows and $\gamma$ indexes columns.
Using \cref{SM_eq:Gamma_g_expansion_fusion} and \cref{SM_eq:Er_Es}, we have
\begin{equation}
    E^r_{\alpha\gamma} 
    = 
    \Re(\omega) 
    [\mathscr{C}_{E A_1}^E]_{\alpha 1}^{\gamma} 
    + 
    i \, 
    \Im(\omega) 
    [\mathscr{C}_{E A_2}^E]_{\alpha 1}^{\gamma} 
    \,.
\end{equation}
Because $A_1$ and $A_2$ are 1D irreps, there is a single term on the right hand side of \cref{SM_eq:ZZ_string_fusion}, so we can directly replace it with the left hand side.
That is,
\begin{align}
    &\begin{tikzpicture}[baseline={(0,.45)}]
        \draw (-1.5, 1.0)--(1.5, 1.0);
        \draw (-1.0,-0.5)--(-1.0,1.5);
        \draw ( 0.0,-0.5)--( 0.0,1.5);
        \draw ( 1.0,-0.5)--( 1.0,1.5);
        \node[mpo](t) at (0,0) {\small $\Xl{r}$};
        \node[mpo](t) at (-1,1) {\small $Z^E$};
        \node[mpo](t) at ( 0,1) {\small $Z^E$};
        \node[mpo](t) at ( 1,1) {\small $Z^E$};
        \node[left]  at (-1.5,1) {$\dots$};
        \node[right] at ( 1.5,1) {$\dots$};
    \end{tikzpicture} 
    =
    \begin{tikzpicture}[baseline={(0,.45)}]
        \draw (-2.5, 0.0)--(1.5, 0.0);
        \draw (-2.0,-0.5)--(-2.0,1.5);
        \draw ( 0.0,-0.5)--( 0.0,1.5);
        \draw ( 1.0,-0.5)--( 1.0,1.5);
        \node[mpo](t) at (0,1)    {\small $\Xl{r}$};
        \node[mpo](t) at (-2,0)   {\small $Z^E$};
        \node[mpo](t) at ( 0,0)   {\small $Z^E$};
        \node[mpo](t) at ( 1,0)   {\small $Z^E$};
        \draw[fill=white, draw=black] (-1,0) circle (.3cm) node {\small $E^r$};        
        \node[left]   at (-2.5,0) {$\dots$};
        \node[right]  at ( 1.5,0) {$\dots$};
    \end{tikzpicture} 
    \nonumber
    \\[2ex]
    &\hspace{3cm}=
    \quad
    \Re(\omega)
    \begin{tikzpicture}[baseline={(0,.45)}]
        \draw (-2.5, 0.0)--(1.5, 0.0);
        \draw (-2.0,-0.5)--(-2.0,1.5);
        \draw ( 0.0,-0.5)--( 0.0,1.5);
        \draw ( 1.0,-0.5)--( 1.0,1.5);
        \node[mpo](t) at (0,1)    {\small $\Xl{r}$};
        \node[mpo](t) at (-2,0)   {\small $Z^E$};
        \node[mpo](t) at ( 0,0)   {\small $Z^E$};
        \node[mpo](t) at ( 1,0)   {\small $Z^E$};
        \draw[fill=white, draw=black] (-1,0) circle (.3cm) node {$\mathbbm{1}$};        
        \node[left]   at (-2.5,0) {$\dots$};
        \node[right]  at ( 1.5,0) {$\dots$};
    \end{tikzpicture} 
    +\quad
    i \, \, 
    \Im(\omega)
    \begin{tikzpicture}[baseline={(0,.45)}]
        \draw (-2.5, 0.0)--(1.5, 0.0);
        \draw (-2.0,-0.5)--(-2.0,1.5);
        \draw ( 0.0,-0.5)--( 0.0,1.5);
        \draw ( 1.0,-0.5)--( 1.0,1.5);
        \node[mpo](t) at (0,1)    {\small $\Xl{r}$};
        \node[mpo](t) at (-2,0)   {\small $Z^E$};
        \node[mpo](t) at ( 0,0)   {\small $Z^E$};
        \node[mpo](t) at ( 1,0)   {\small $Z^E$};
        \draw[fill=white, draw=black] (-1,0) circle (.3cm) node {\small $\sigma^z$};        
        \node[left]   at (-2.5,0) {$\dots$};
        \node[right]  at ( 1.5,0) {$\dots$};
    \end{tikzpicture} 
    \nonumber
    \\
    &\hspace{3cm}=
    \quad
    \Re(\omega)
    \begin{tikzpicture}[baseline={(0,.45)}]
        \draw (-1.5, 0.0)--(1.5, 0.0);
        \draw (-1.0,-0.5)--(-1.0,1.5);
        \draw ( 0.0,-0.5)--( 0.0,1.5);
        \draw ( 1.0,-0.5)--( 1.0,1.5);
        \node[mpo](t) at (0,1)    {\small $\Xl{r}$};
        \node[mpo](t) at (-1,0)   {\small $Z^E$};
        \node[mpo](t) at ( 0,0)   {\small $Z^E$};
        \node[mpo](t) at ( 1,0)   {\small $Z^E$};
        \node[left]   at (-1.5,0) {$\dots$};
        \node[right]  at ( 1.5,0) {$\dots$};
    \end{tikzpicture} 
    +\quad
    i \, \, \Im(\omega)
    \begin{tikzpicture}[baseline={(0,.45)}]
        \draw (-2.5, 0.0)--(0.5, 0.0);
        \draw (-2.0,-0.5)--(-2.0,1.5);
        \draw (-1.0,-0.5)--(-1.0,1.5);
        \draw ( 0.0,-0.5)--( 0.0,1.5);
        \node[mpo](t) at (-2,0)   {\small $Z^E$};
        \node[mpo](t) at (-1,0)   {\small $Z^E$};
        \node[mpo](t) at ( 0,0)   {\small $Z^E$};
        \node[mpo](t) at (-1,1)   {\small $\Xl{r}$};
        \node[mpo](t) at (-2,1)   {\small $Z^{A_2}$};
        \node[left]   at (-2.5,0) {$\dots$};
        \node[left]   at (-2.5,1) {$\dots$};
        \node[right]  at (0.5,0) {$\dots$};
    \end{tikzpicture} 
    \label{SM_eq:XRE-string}
\end{align}
where in the final line we pulled out a half-infinite $A_2$ string, which inserts a $\sigma^z$ according to the $E\otimes A_2\to E$ Clebsch-Gordan coefficients.\footnote{
    We could also have pulled the $\smash{Z^{A_2}}$ string out on the right side, with the only downside being that it would overlap with the $\smash{\Xl{r}}$. This overlap is unambiguous because $\smash{[\Xl{r},Z^{A_2}]=0}$. 
}
In other words, we have the relation
\begin{equation}
    R_E \Xl{r}_j =  \left(\Re(\omega) \Xl{r}_j +i \, \Im(\omega)\prod_{i < j} Z_i^{A_2} \Xl{r}_j\right) R_E 
    \, ,
\end{equation}
or, schematically,
\begin{equation}
    \begin{tikzpicture}[baseline={(0,-0.1)}]
        \draw[thick] (0,0)--(2,0);
        \filldraw (1,-0.5) circle (.5mm);
        \node[left]  at (1,-0.5) {$r$};
        \node[right] at (2,0) {$E$};
    \end{tikzpicture}
    \quad
    =
    \quad
    \Re(\omega)
    \quad
    \begin{tikzpicture}[baseline={(0,-0.1)}]
        \draw[thick] (0,0)--(2,0);
        \filldraw (1, 0.5) circle (.5mm);
        \node[left]  at (1, 0.5) {$r$};
        \node[right] at (2,0) {$E$};
    \end{tikzpicture}
    \quad
    +
    \quad
    i \,
    \Im(\omega)
    \quad
    \begin{tikzpicture}[baseline={(0,-0.1)}]
        \draw[thick] (0,0)--(2,0);
        \draw[decorate, decoration={snake, amplitude=.5mm, segment length=2mm}] (0.7,0.6)--(2,0.6);
        \filldraw (0.7,0.6) circle (.5mm);
        \node[left]  at (0.7,0.6) {$r$};
        \node[right] at (2,0.0)   {$E$};
        \node[right] at (2,0.6)   {$A_2$};
    \end{tikzpicture}
    \,.
\end{equation}
Next we will do the same for $E^s$. 
For simplicity we consider pulling the operator $\Xr{s}$ through the $R_E$ MPO (note that $s = s^{-1}$), 
\begin{equation}
    \begin{tikzpicture}[baseline={(0,.45)}]
        \draw (-1.5, 1.0)--(1.5, 1.0);
        \draw (-1.0,-0.5)--(-1.0,1.5);
        \draw ( 0.0,-0.5)--( 0.0,1.5);
        \draw ( 1.0,-0.5)--( 1.0,1.5);
        \node[mpo](t) at (-1,0) {\small $\Xr{s}$};
        \node[mpo](t) at (-1,1) {\small $Z^E$};
        \node[mpo](t) at ( 0,1) {\small $Z^E$};
        \node[mpo](t) at ( 1,1) {\small $Z^E$};
        \node[left]  at (-1.5,1) {$\dots$};
        \node[right] at ( 1.5,1) {$\dots$};
    \end{tikzpicture} 
    =
    \begin{tikzpicture}[baseline={(0,.45)}]
        \draw (-2.5, 0.0)--(1.5, 0.0);
        \draw (-2.0,-0.5)--(-2.0,1.5);
        \draw ( 0.0,-0.5)--( 0.0,1.5);
        \draw ( 1.0,-0.5)--( 1.0,1.5);
        \node[mpo](t) at (-2,1)    {\small $\Xr{s}$};
        \node[mpo](t) at (-2,0)   {\small $Z^E$};
        \node[mpo](t) at ( 0,0)   {\small $Z^E$};
        \node[mpo](t) at ( 1,0)   {\small $Z^E$};
        \draw[fill=white, draw=black] (-1,0) circle (.3cm) node {\small $E^s$};        
        \node[left]   at (-2.5,0) {$\dots$};
        \node[right]  at ( 1.5,0) {$\dots$};
    \end{tikzpicture} 
    \label{SM_eq:XsE_EXs}
\end{equation}
From \cref{SM_eq:Er_Es},
\begin{equation}
    E^s_{\alpha\gamma} 
    = 
    \sigma^x_{\alpha\gamma} 
    = 
    \sigma^+_{\alpha\gamma} + \sigma^-_{\alpha\gamma} = 
    [\mathscr{C}_{E E}^E]_{\alpha 1}^{\gamma} 
    + 
    [\mathscr{C}_{E E}^E]_{\alpha 2}^{\gamma} 
    \,.
\end{equation}
From this it must be that the $E^s$ defect is inserted by a half-infinite $E$ string with the open end contracted with $\ket{1}+\ket{2}\equiv\sqrt{2}\ket{+}$, which we denote
\begin{equation}
    \sqrt{2}
    \quad
    \begin{tikzpicture}[baseline={(0,-0.1)}]
        \def\l{5mm}
        \draw (0.3,0.0)--(2.6,0.0);
        \draw ( 1.0,-0.5)--( 1.0,0.5);
        \draw ( 2.0,-0.5)--( 2.0,0.5);
        \draw[fill=white,draw=black] (0.1,0) -- ++(30:\l) -- ++(-90:\l) -- cycle;
        \node at (0.4,0) {\small $+$};
        \node[mpo](t) at (1.0,0.0) {\small $Z^E$};
        \node[mpo](t) at (2.0,0.0) {\small $Z^E$};
        \node[right] at (2.6,0.0) {\small $\dots$};
    \end{tikzpicture}
    \equiv 
    \sum_{\beta=1}^2
    \,
    \sqrt{2}
    \braket{+ \vert \beta}
    \begin{tikzpicture}[baseline={(0,-0.1)}]
        \draw (0.3,0.0)--(2.6,0.0);
        \draw ( 1.0,-0.5)--( 1.0,0.5);
        \draw ( 2.0,-0.5)--( 2.0,0.5);
        \filldraw (0.3,0) circle (.5mm);
        \node[above] at (0.3,0) {\small $\beta$};
        \node[mpo](t) at (1.0,0.0) {\small $E$};
        \node[mpo](t) at (2.0,0.0) {\small $E$};
        \node[right] at (2.6,0.0) {\small $\dots$};
    \end{tikzpicture}
    \equiv 
    \sum_{\beta=1}^2
    \,
    \begin{tikzpicture}[baseline={(0,-0.1)}]
        \draw (0.3,0.0)--(2.6,0.0);
        \draw ( 1.0,-0.5)--( 1.0,0.5);
        \draw ( 2.0,-0.5)--( 2.0,0.5);
        \filldraw (0.3,0) circle (.5mm);
        \node[above] at (0.3,0) {\small $\beta$};
        \node[mpo](t) at (1.0,0.0) {\small $E$};
        \node[mpo](t) at (2.0,0.0) {\small $E$};
        \node[right] at (2.6,0.0) {\small $\dots$};
    \end{tikzpicture}
\end{equation}
However, since there are multiple terms on the right hand side of \cref{SM_eq:ZZ_string_fusion} for $a=b=E$, we cannot simply pull out this string operator. 
Multiplying such an open $E$ string with the $R_E$ operator, using the Clebsch-Gordan coefficients
\begin{equation}
    [\mathscr{C}_{E E}^{A_1}]_{\alpha \beta}^{1} 
    = 
    \frac{1}{\sqrt{2}}
    \begin{pmatrix*}
        0 & 1 \\ 1 & 0
    \end{pmatrix*}
    =
    \frac{1}{\sqrt{2}}
    \sigma^x
    ,
    \quad
    [\mathscr{C}_{E E}^{A_2}]_{\alpha \beta}^{1} 
    = 
    \frac{1}{\sqrt{2}}
    \begin{pmatrix*}
        0 & 1 \\ -1 & 0
    \end{pmatrix*}
    =
    \frac{1}{\sqrt{2}}
    i\sigma^y
    \,,
    \label{SM_eq:E_E_fusion}
\end{equation}
we can write out \cref{SM_eq:ZZ_string_fusion},
\begin{align}
    \sqrt{2}
    \begin{tikzpicture}[baseline={(0,-0.1)}]
        \def\l{5mm}
        \draw (0.4,0.0)--(2.6,0.0);
        \draw (1.4,0.8)--(2.6,0.8);
        \draw ( 1.0,-0.5)--( 1.0,1.3);
        \draw ( 2.0,-0.5)--( 2.0,1.3);
        \node[mpo](t) at (2.0,0.8) {\small $Z^E$};
        \node[mpo](t) at (1.0,0.0) {\small $Z^E$};
        \node[mpo](t) at (2.0,0.0) {\small $Z^E$};
        \draw[fill=white,draw=black] (1.1,0.8) -- ++(30:\l) -- ++(-90:\l) -- cycle;
        \node at (1.4,0.8) {\small $+$};
        \node[left]  at (0.4,0.0) {\small $\dots$};
        \node[right] at (2.6,0.0) {\small $\dots$};
        \node[right] at (2.6,0.8) {\small $\dots$};
    \end{tikzpicture}
    &=
    \sum_{\beta}
    \Bigg(
    \begin{tikzpicture}[baseline={(0,-0.1)}]
        \draw (-1.5,0)--(.4,0);
        \draw (-0.9,-0.6)--(-0.9,0.6);
        \draw ( 1.1,-0.6)--( 1.1,0.6);
        \node[mpo](t) at (-0.9,0) {\small $Z^E$};   
        \node[mpo](t) at ( 1.1,0) {\small $Z^{A_1}$};
        \draw (-.1,0)--(-.1,.5);
        \filldraw (-.1,0.5) circle (.5mm);
        \node[right] at (-.1,.5)  {\small $\,1$};
        \node[below] at (.4,-.05) {\small $\beta$};
        \filldraw (.4,0) circle (.5mm);
        \node[below] at (.4,-.05) {\small $\beta$};
        \draw[fill=white, draw=black] (-.1,0) circle (.35cm) node {\small \scalebox{0.8}{$\mathscr{C}_{EE}^{A_1}$}};
        \node[right] at (1.4,0.0) {\small $\dots$};
    \end{tikzpicture}
    \,\,
    +
    \quad
    \begin{tikzpicture}[baseline={(0,-0.1)}]
        \draw (-1.5,0)--(.4,0);
        \draw (-0.9,-0.6)--(-0.9,0.6);
        \draw ( 1.1,-0.6)--( 1.1,0.6);
        \node[mpo](t) at (-0.9,0) {\small $Z^E$};   
        \node[mpo](t) at ( 1.1,0) {\small $Z^{A_2}$};
        \draw (-.1,0)--(-.1,.5);
        \filldraw (-.1,0.5) circle (.5mm);
        \node[right] at (-.1,.5)  {\small $\,1$};
        \filldraw (.4,0) circle (.5mm);
        \node[below] at (.4,-.05) {\small $\beta$};
        \draw[fill=white, draw=black] (-.1,0) circle (.35cm) node {\small \scalebox{0.8}{$\mathscr{C}_{EE}^{A_2}$}};
        \node[right] at (1.4,0.0) {\small $\dots$};
    \end{tikzpicture}
    \,\,
    +
    \quad
    \begin{tikzpicture}[baseline={(0,-0.1)}]
        \draw (-1.5,0)--(.4,0);
        \draw (1.1,0)--(1.6,0);
        \draw (-0.1,0.33) to[out=90, in=180] (.8,0);
        \draw (-0.9,-0.6)--(-0.9,0.6);
        \draw ( 1.1,-0.6)--( 1.1,0.6);
        \node[mpo](t) at (-0.9,0) {\small $Z^E$};   
        \node[mpo](t) at ( 1.1,0) {\small $Z^E$};
        \filldraw (.4,0) circle (.5mm);
        \node[below] at (.4,-.05) {\small $\beta$};
        \draw[fill=white, draw=black] (-.1,0) circle (.35cm) node {\small \scalebox{0.8}{$\mathscr{C}_{EE}^{E}$}};
    \end{tikzpicture}
    \Bigg)
    \nonumber
    \\
    &=
    \begin{tikzpicture}[baseline={(0,-0.1)}]
        \def\l{5mm}
        \draw (-1.5,0)--(0,0);
        \draw (-0.9,-0.6)--(-0.9,0.6);
        \draw ( 0.5,-0.6)--( 0.5,0.6);
        \node[mpo](t) at (-0.9,0) {\small $Z^E$};   
        \node[mpo](t) at ( 0.5,0) {\small $Z^{A_1}$};
        \draw[fill=white,draw=black] (0,0) -- ++(150:\l) -- ++(-90:\l) -- cycle;
        \node at (-.27,0) {\small $+$};
        \node[left]  at (-1.5,0.0) {\small $\dots$};
        \node[right] at (0.8,0.0) {\small $\dots$};
    \end{tikzpicture}
    \,\,
    +
    \,\,
    \begin{tikzpicture}[baseline={(0,-0.1)}]
        \def\l{5mm}
        \draw (-1.5,0)--(0,0);
        \draw (-0.9,-0.6)--(-0.9,0.6);
        \draw ( 0.5,-0.6)--( 0.5,0.6);
        \node[mpo](t) at (-0.9,0) {\small $Z^E$};   
        \node[mpo](t) at ( 0.5,0) {\small $Z^{A_2}$};
        \draw[fill=white,draw=black] (0,0) -- ++(150:\l) -- ++(-90:\l) -- cycle;
        \node at (-.27,-0.02) {\small $-$};
        \node[left]  at (-1.5,0.0) {\small $\dots$};
        \node[right] at (0.8,0.0) {\small $\dots$};
    \end{tikzpicture}
    \,\,
    +
    \,\,
    \begin{tikzpicture}[baseline={(0,-0.1)}]
        \draw (-1.5,0)--(1.6,0);
        \draw (-0.9,-0.6)--(-0.9,0.6);
        \draw ( 1.1,-0.6)--( 1.1,0.6);
        \node[mpo](t) at (-0.9,0) {\small $Z^E$};   
        \node[mpo](t) at ( 1.1,0) {\small $Z^E$};
        \draw[fill=white, draw=black] (.1,0) circle (.35cm) node {\small $\sigma^x$};
    \end{tikzpicture}
    \nonumber
    \\[2ex]
    &=
    \begin{tikzpicture}[baseline={(0,-0.1)}]
        \def\l{5mm}
        \draw (-1.5,0)--(0,0);
        \draw (-0.9,-0.6)--(-0.9,0.6);
        \draw ( 0.5,-0.6)--( 0.5,0.6);
        \node[mpo](t) at (-0.9,0) {\small $Z^E$};  
        \node[mpo](t) at ( 0.5,0) {\small $Z^{A_1}$};
        \draw[fill=white,draw=black] (0,0) -- ++(150:\l) -- ++(-90:\l) -- cycle;
        \node at (-.27,0) {\small $+$};
        \node[left]  at (-1.5,0.0) {\small $\dots$};
        \node[right] at (0.8,0.0) {\small $\dots$};
    \end{tikzpicture}
    \,\,
    +
    \,\,
    \begin{tikzpicture}[baseline={(0,-0.1)}]
        \def\l{5mm}
        \draw (-1.5,0)--(0,0);
        \draw (-0.9,-0.6)--(-0.9,0.6);
        \draw ( 0.5,-0.6)--( 0.5,0.6);
        \node[mpo](t) at (-0.9,0) {\small $Z^E$};   
        \node[mpo](t) at ( 0.5,0) {\small $Z^{A_2}$};
        \draw[fill=white,draw=black] (0,0) -- ++(150:\l) -- ++(-90:\l) -- cycle;
        \node at (-.27,-0.02) {\small $-$};
        \node[left]  at (-1.5,0.0) {\small $\dots$};
        \node[right] at (0.8,0.0) {\small $\dots$};
    \end{tikzpicture}
    \,\,
    +
    \,\,
    \begin{tikzpicture}[baseline={(0,-0.1)}]
        \draw (-1.5,0)--(1.6,0);
        \draw (-0.9,-0.6)--(-0.9,0.6);
        \draw ( 1.1,-0.6)--( 1.1,0.6);
        \node[mpo](t) at (-0.9,0) {\small $Z^E$};   
        \node[mpo](t) at ( 1.1,0) {\small $Z^E$};
        \draw[fill=white, draw=black] (.1,0) circle (.35cm) node {\small $E^s$};
    \end{tikzpicture}
    \,.
\end{align}
In the second line we summed \cref{SM_eq:E_E_fusion} over $\beta$ to obtain the vectors $\ket{\pm} = (\ket{1} \pm \ket{2})/\sqrt{2}$. 
In the last line we identified $\sigma^x$ as $E^s$.
With this we can write \cref{SM_eq:XsE_EXs} as \begin{equation}
    \begin{tikzpicture}[baseline={(0,.45)}]
        \draw (-1.5, 1.0)--(1.5, 1.0);
        \draw (-1.0,-0.5)--(-1.0,1.5);
        \draw ( 0.0,-0.5)--( 0.0,1.5);
        \draw ( 1.0,-0.5)--( 1.0,1.5);
        \node[mpo](t) at (-1,0) {\small $\Xr{s}$};
        \node[mpo](t) at (-1,1) {\small $Z^E$};
        \node[mpo](t) at ( 0,1) {\small $Z^E$};
        \node[mpo](t) at ( 1,1) {\small $Z^E$};
        \node[left]  at (-1.5,1) {$\dots$};
        \node[right] at ( 1.5,1) {$\dots$};
    \end{tikzpicture} 
    =
    \sqrt{2}
    \begin{tikzpicture}[baseline={(0,0.3)}]
        \def\l{5mm}
        \draw (0.0,0.0)--(2.6,0.0);
        \draw (1.4,0.8)--(2.6,0.8);
        \draw ( 0.5,-0.5)--( 0.5,1.3);
        \draw ( 2.0,-0.5)--( 2.0,1.3);
        \node[mpo](t) at (0.5,0.8) {\small $\Xr{s}$};
        \node[mpo](t) at (2.0,0.8) {\small $Z^E$};
        \node[mpo](t) at (0.5,0.0) {\small $Z^E$};
        \node[mpo](t) at (2.0,0.0) {\small $Z^E$};
        \draw[fill=white,draw=black] (1.1,0.8) -- ++(30:\l) -- ++(-90:\l) -- cycle;
        \node at (1.4,0.8) {\small $+$};
        \node[left]  at (0.0,0.0) {\small $\dots$};
        \node[right] at (2.6,0.0) {\small $\dots$};
        \node[right] at (2.6,0.8) {\small $\dots$};
    \end{tikzpicture}
    -
    \begin{tikzpicture}[baseline={(0,0.3)}]
        \def\l{5mm}
        \draw (-1.5,0)--(0,0);
        \draw (-0.9,-0.6)--(-0.9,1.3);
        \draw ( 0.5,-0.6)--( 0.5,1.3);
        \node[mpo](t) at (-0.9,0.8) {\small $\Xr{s}$};
        \node[mpo](t) at (-0.9,0) {\small $Z^E$};          %
        \node[mpo](t) at (0.5,0) {\small $Z^{A_1}$};  
        \draw[fill=white,draw=black] (0,0) -- ++(150:\l) -- ++(-90:\l) -- cycle;
        \node at (-.27,0) {\small $+$};
        \node[left]  at (-1.5,0.0) {\small $\dots$};
        \node[right] at (0.8,0.0) {\small $\dots$};
    \end{tikzpicture}
    \,\,
    -
    \,\,
    \begin{tikzpicture}[baseline={(0,0.3)}]
        \def\l{5mm}
        \draw (-1.5,0)--(0,0);
        \draw (-0.9,-0.6)--(-0.9,1.3);
        \draw ( 0.5,-0.6)--( 0.5,1.3);
        \node[mpo](t) at (-0.9,0.8) {\small $\Xr{s}$};
        \node[mpo](t) at (-0.9,0) {\small $Z^E$};   
        \node[mpo](t) at ( 0.5,0) {\small $Z^{A_2}$};
        \draw[fill=white,draw=black] (0,0) -- ++(150:\l) -- ++(-90:\l) -- cycle;
        \node at (-.27,-0.02) {\small $-$};
        \node[left]  at (-1.5,0.0) {\small $\dots$};
        \node[right] at (0.8,0.0) {\small $\dots$};
    \end{tikzpicture}
\end{equation}
Schematically, this equation reads
\begin{equation}
    \begin{tikzpicture}[baseline={(0,-0.1)}]
        \draw[thick] (0,0)--(2,0);
        \filldraw (0.7,-0.5) circle (.5mm);
        \node[left]  at (0.7,-0.5) {$s$};
        \node[left]  at (0,0.0) {$E$};
        \node[right] at (2,0) {$E$};
    \end{tikzpicture}
    \quad
    =
    \quad
    \sqrt{2}
    \quad
    \begin{tikzpicture}[baseline={(0,-0.1)}]
        \draw[thick] (0,0)--(2,0);
        \draw[thick] (0.7,0.6)--(2,0.6);
        \draw[dashed] (0,0.6)--(0.7,0.6);
        \filldraw (0.7,0.6) circle (.5mm);
        \node[above] at (0.7,0.6) {$s$};
        \node[below] at (0.7,0.6) {$+$};
        \node[left]  at (0,0.0) {$E$};
        \node[left]  at (0,0.6) {$A_1$};
        \node[right] at (2,0.0) {$E$};
        \node[right] at (2,0.6) {$E$};
    \end{tikzpicture}
    \,\,
    -
    \,\,
    \begin{tikzpicture}[baseline={(0,-0.1)}]
        \draw[thick]  (0,0)--(1,0);
        \draw[dashed] (1,0)--(2,0);
        \filldraw (1,0) circle (.5mm);
        \node[above] at (1,.05) {$s$};
        \node[below] at (1,0) {$+$};
        \node[left]  at (0,0) {$E$};
        \node[right] at (2,0) {$A_1$};
    \end{tikzpicture}
    \,\,
    -
    \,\,
    \begin{tikzpicture}[baseline={(0,-0.1)}]
        \draw[thick]  (0,0)--(1,0);
        \draw[decorate, decoration={snake, amplitude=.5mm, segment length=2mm}] (1,0)--(2,0);
        \filldraw (1,0) circle (.5mm);
        \node[above] at (1, .05) {$s$};
        \node[below] at (1,-.05) {$-$};
        \node[left]  at (0,0) {$E$};
        \node[right] at (2,0) {$A_2$};
    \end{tikzpicture}
\end{equation}
where each line represents a string of $Z^\Gamma$ operators (note that $Z^{A_1}$ is the identity).

\section{Relations Between Local and String Order Parameters}
First, we define the Hermitian operators
\begin{align}
    \Ol{+}_i 
    &= 
    \Xl{r}_i + {\Xl{r}_i}^{\dagger} 
    = 
    \Xl{r}_i + \Xl{r^{-1}}_i
    \qquad
    \Ol{-}_i 
    = 
    \frac{1}{i}\left(\Xl{r}_i - {\Xl{r}_i}^{\dagger} \right)
    = 
    \frac{1}{i}\left(\Xl{r}_i - \Xl{r^{-1}}_i\right)
    \\[1ex]
    \Or{+}_i 
    &= 
    \Xr{r}_i + {\Xr{r}_i}^{\dagger} 
    = 
    \Xr{r}_i + \Xr{r^{-1}}_i
    \qquad
    \Or{-}_i 
    = 
    \frac{1}{i}\left(\Xr{r}_i - {\Xr{r}_i}^{\dagger} \right)
    = 
    \frac{1}{i}\left(\Xr{r}_i - \Xr{r^{-1}}_i\right)
\end{align}
Pulling a single $\Ol{\pm}$ through the $R_E$ MPO corresponds to adding/subtracting \cref{SM_eq:XRE-string} to itself with $r \to r^{-1}$ and $\omega \to \omega^*$, so we obtain
\begin{equation}
    \begin{tikzpicture}[baseline={(0,.45)}]
        \draw (-1.5, 1.0)--(1.5, 1.0);
        \draw (-1.0,-0.5)--(-1.0,1.5);
        \draw ( 0.0,-0.5)--( 0.0,1.5);
        \draw ( 1.0,-0.5)--( 1.0,1.5);
        \node[mpo](t) at (0,0) {$\Ol{\pm}$};
        \node[mpo](t) at (-1,1) {\small $Z^E$};
        \node[mpo](t) at ( 0,1) {\small $Z^E$};
        \node[mpo](t) at ( 1,1) {\small $Z^E$};
        \node[left]  at (-1.5,1) {$\dots$};
        \node[right] at ( 1.5,1) {$\dots$};
    \end{tikzpicture} 
    =
    \,
    \Re(\omega)
    \begin{tikzpicture}[baseline={(0,.45)}]
        \draw (-1.5, 0.0)--(1.5, 0.0);
        \draw (-1.0,-0.5)--(-1.0,1.5);
        \draw ( 0.0,-0.5)--( 0.0,1.5);
        \draw ( 1.0,-0.5)--( 1.0,1.5);
        \node[mpo](t) at (0,1)    {$\Ol{\pm}$};
        \node[mpo](t) at (-1,0)   {\small $Z^E$};
        \node[mpo](t) at ( 0,0)   {\small $Z^E$};
        \node[mpo](t) at ( 1,0)   {\small $Z^E$};
        \node[left]   at (-1.5,0) {$\dots$};
        \node[right]  at ( 1.5,0) {$\dots$};
    \end{tikzpicture} 
    \mp
    \Im(\omega)
    \begin{tikzpicture}[baseline={(0,.45)}]
        \draw (-2.5, 0.0)--(1.5, 0.0);
        \draw (-2.0,-0.5)--(-2.0,1.5);
        \draw ( 0.0,-0.5)--( 0.0,1.5);
        \draw ( 1.0,-0.5)--( 1.0,1.5);
        \node[mpo](t) at (0,1)    {$\Ol{\mp}$};
        \node[mpo](t) at (-2,0)   {\small $Z^E$};
        \node[mpo](t) at ( 0,0)   {\small $Z^E$};
        \node[mpo](t) at ( 1,0)   {\small $Z^E$};
        \draw[fill=white, draw=black] (-1,0) circle (.3cm) node {\small $\sigma^z$};        
        \node[left]   at (-2.5,0) {$\dots$};
        \node[right]  at ( 1.5,0) {$\dots$};
    \end{tikzpicture} 
\end{equation}
Note that in the last term $\mathcal{O}^{\pm}$ changed to $\mathcal{O}^{\mp}$ and there is no imaginary unit.
Similarly, for $\Or{\pm}$ we find
\begin{equation}
    \begin{tikzpicture}[baseline={(0,.45)}]
        \draw (-1.5, 1.0)--(1.5, 1.0);
        \draw (-1.0,-0.5)--(-1.0,1.5);
        \draw ( 0.0,-0.5)--( 0.0,1.5);
        \draw ( 1.0,-0.5)--( 1.0,1.5);
        \node[mpo](t) at (0,0) {$\Or{\pm}$};
        \node[mpo](t) at (-1,1) {\small $Z^E$};
        \node[mpo](t) at ( 0,1) {\small $Z^E$};
        \node[mpo](t) at ( 1,1) {\small $Z^E$};
        \node[left]  at (-1.5,1) {$\dots$};
        \node[right] at ( 1.5,1) {$\dots$};
    \end{tikzpicture} 
    =
    \,
    \Re(\omega)
    \begin{tikzpicture}[baseline={(0,.45)}]
        \draw (-1.5, 0.0)--(1.5, 0.0);
        \draw (-1.0,-0.5)--(-1.0,1.5);
        \draw ( 0.0,-0.5)--( 0.0,1.5);
        \draw ( 1.0,-0.5)--( 1.0,1.5);
        \node[mpo](t) at (0,1)    {$\Or{\pm}$};
        \node[mpo](t) at (-1,0)   {\small $Z^E$};
        \node[mpo](t) at ( 0,0)   {\small $Z^E$};
        \node[mpo](t) at ( 1,0)   {\small $Z^E$};
        \node[left]   at (-1.5,0) {$\dots$};
        \node[right]  at ( 1.5,0) {$\dots$};
    \end{tikzpicture} 
    \pm
    \Im(\omega)
    \begin{tikzpicture}[baseline={(0,.45)}]
        \draw (-2.5, 0.0)--(1.5, 0.0);
        \draw (-2.0,-0.5)--(-2.0,1.5);
        \draw (-1.0,-0.5)--(-1.0,1.5);
        \draw ( 1.0,-0.5)--( 1.0,1.5);
        \node[mpo](t) at (-1,1)   {$\Or{\mp}$};
        \node[mpo](t) at (-2,0)   {\small $Z^E$};
        \node[mpo](t) at (-1,0)   {\small $Z^E$};
        \node[mpo](t) at ( 1,0)   {\small $Z^E$};
        \draw[fill=white, draw=black] (0,0) circle (.3cm) node {\small $\sigma^z$};        
        \node[left]   at (-2.5,0) {$\dots$};
        \node[right]  at ( 1.5,0) {$\dots$};
    \end{tikzpicture} 
\end{equation}
Now we want to examine what happens when we pull the product $\Or{\pm}_i \Ol{\pm}_j$ through the $R_E$ MPO with $i<j$. Diagrammatically, we find 
\begin{align}
    &\begin{tikzpicture}[baseline={(0,.45)}]
        \draw (-1.5, 1.0)--(1.5, 1.0);
        \draw (-1.0,-0.5)--(-1.0,1.5);
        \draw ( 1.0,-0.5)--( 1.0,1.5);
        \node[mpo](t) at (-1,0) {$\Or{\pm}$};
        \node[mpo](t) at ( 1,0) {$\Ol{\pm}$};
        \node[mpo](t) at (-1,1) {\small $Z^E$};
        \node[mpo](t) at ( 1,1) {\small $Z^E$};
        \node[left]  at (-1.5,1) {$\dots$};
        \node[right] at ( 1.5,1) {$\dots$};
        \node[fill=white] at (0,1) {$\dots$};
        \node[below] at (-1,-.5) {$i$};
        \node[below] at ( 1,-.5) {$j$};
    \end{tikzpicture} 
    =
    \cdots
    \left(
    \Re(\omega)
    \begin{tikzpicture}[baseline={(0,.45)}]
        \draw (-.5, 0.0)--(0.5, 0.0);
        \draw (0.0,-0.5)--(0.0, 1.5);
        \node[mpo](t) at (0,1) {$\Or{\pm}$};
        \node[mpo](t) at (0,0) {\small $Z^E$};
    \end{tikzpicture} 
    \pm
    \Im(\omega)
    \begin{tikzpicture}[baseline={(0,.45)}]
        \draw (-1.5, 0.0)--( 0.5, 0.0);
        \draw (-1.0,-0.5)--(-1.0, 1.5);
        \draw[fill=white, draw=black] (0,0) circle (.3cm) node {\small $\sigma^z$};        
        \node[mpo](t) at (-1,1) {$\Or{\mp}$};
        \node[mpo](t) at (-1,0) {\small $Z^E$};
    \end{tikzpicture} 
    \right)_i
    \cdots
    \left(
    \Re(\omega)
    \begin{tikzpicture}[baseline={(0,.45)}]
        \draw (-.5, 0.0)--(0.5, 0.0);
        \draw (0.0,-0.5)--(0.0, 1.5);
        \node[mpo](t) at (0,1) {$\Ol{\pm}$};
        \node[mpo](t) at (0,0) {\small $Z^E$};
    \end{tikzpicture} 
    \mp
    \Im(\omega)
    \begin{tikzpicture}[baseline={(0,.45)}]
        \draw (-1.5, 0.0)--(0.5, 0.0);
        \draw (0.0,-0.5)--(0.0, 1.5);
        \draw[fill=white, draw=black] (-1,0) circle (.3cm) node {\small $\sigma^z$};        
        \node[mpo](t) at (0,1) {$\Ol{\mp}$};
        \node[mpo](t) at (0,0) {\small $Z^E$};
    \end{tikzpicture} 
    \right)_j
    \cdots
    \nonumber
    \\[2ex]
    &\qquad=\quad
    \Re(\omega)^2
    \left(
    \begin{tikzpicture}[baseline={(0,.45)}]
        \draw (-1.5, 0.0)--(1.5, 0.0);
        \draw (-1.0,-0.5)--(-1.0,1.5);
        \draw ( 1.0,-0.5)--( 1.0,1.5);
        \node[mpo](t) at (-1,1) {$\Or{\pm}$};
        \node[mpo](t) at ( 1,1) {$\Ol{\pm}$};
        \node[mpo](t) at (-1,0) {\small $Z^E$};
        \node[mpo](t) at ( 1,0) {\small $Z^E$};
        \node[left]  at (-1.5,0) {$\dots$};
        \node[right] at ( 1.5,0) {$\dots$};
        \node[fill=white] at (0,0) {$\dots$};
        \node[below] at (-1,-.5) {$i$};
        \node[below] at ( 1,-.5) {$j$};
    \end{tikzpicture} 
    \right)
    -
    \Im(\omega)^2
    \left(
    \begin{tikzpicture}[baseline={(0,.45)}]
        \draw (-2.5, 0.0)--(2.5, 0.0);
        \draw (-2.0,-0.5)--(-2.0,1.5);
        \draw (-1.0,-0.5)--(-1.0,1.5);
        \draw ( 1.0,-0.5)--( 1.0,1.5);
        \draw ( 2.0,-0.5)--( 2.0,1.5);
        \node[mpo](t) at (-2,1) {$\Or{\mp}$};
        \node[mpo](t) at ( 2,1) {$\Ol{\mp}$};
        \node[mpo](t) at (-2,0) {\small $Z^E$};
        \node[mpo](t) at (-1,0) {\small $Z^E$};
        \node[mpo](t) at ( 1,0) {\small $Z^E$};
        \node[mpo](t) at ( 2,0) {\small $Z^E$};
        \node[mpo](t) at (-1,1) {\small $Z^{A_2}$};
        \node[mpo](t) at ( 1,1) {\small $Z^{A_2}$};
        \node[left]  at (-2.5,0) {$\dots$};
        \node[right] at ( 2.5,0) {$\dots$};
        \node[fill=white] at (0,0) {$\dots$};
        \node[fill=white] at (0,1) {$\dots$};
        \node[below] at (-2,-.5) {$i$};
        \node[below] at ( 2,-.5) {$j$};
    \end{tikzpicture} 
    \right)
    \nonumber
    \\[2ex]
    &\qquad \qquad
    \mp
    \Re(\omega)\Im(\omega)
    \left(
    \begin{tikzpicture}[baseline={(0,.45)}]
        \draw (-.5, 0.0)--(0.5, 0.0);
        \draw (0.0,-0.5)--(0.0, 1.5);
        \node[mpo](t) at (0,1) {$\Or{\pm}$};
        \node[mpo](t) at (0,0) {\small $Z^E$};
        \node[right] at (.5,0) {$\dots$};
    \end{tikzpicture} 
    \begin{tikzpicture}[baseline={(0,.45)}]
        \draw (-1.5, 0.0)--(0.5, 0.0);
        \draw (0.0,-0.5)--(0.0, 1.5);
        \draw[fill=white, draw=black] (-1,0) circle (.3cm) node {\small $\sigma^z$};        
        \node[mpo](t) at (0,1) {$\Ol{\mp}$};
        \node[mpo](t) at (0,0) {\small $Z^E$};
    \end{tikzpicture} 
    \quad
    -
    \quad
    \begin{tikzpicture}[baseline={(0,.45)}]
        \draw (-1.5, 0.0)--( 0.5, 0.0);
        \draw (-1.0,-0.5)--(-1.0, 1.5);
        \draw[fill=white, draw=black] (0,0) circle (.3cm) node {\small $\sigma^z$};        
        \node[mpo](t) at (-1,1) {$\Or{\mp}$};
        \node[mpo](t) at (-1,0) {\small $Z^E$};
        \node[right] at (.5,0) {$\dots$};
    \end{tikzpicture} 
    \begin{tikzpicture}[baseline={(0,.45)}]
        \draw (-.5, 0.0)--(0.5, 0.0);
        \draw (0.0,-0.5)--(0.0, 1.5);
        \node[mpo](t) at (0,1) {$\Ol{\pm}$};
        \node[mpo](t) at (0,0) {\small $Z^E$};
    \end{tikzpicture} 
    \right)
\end{align}
Using this, we can relate the two-point correlators of the $\mathcal{O}$ operators to the string order parameter. Working in the $h=0$ limit, we have on the one hand
\begin{equation}
    \bra{\bm E} R_E \Or{\pm}_i \Ol{\pm}_j \ket{\bm A_1} = \bra{\bm A_1} \Or{\pm}_i \Ol{\pm}_j \ket{\bm A_1}
    \label{SM_eq:relation_1},
\end{equation}
where we used the identity $R_E \ket{\bm E} = \sum_{\Gamma} \ket{\bm \Gamma}$ and the fact that local operators have vanishing matrix elements when sandwiched between different SSB ground states. On the other hand, we can pull the $R_E$ through to the right, yielding four terms
\begin{align}
    \bra{\bm E} R_E \Or{\pm}_i \Ol{\pm}_j \ket{\bm A_1} 
    &= 
    \Re(\omega)^2 
    \bra{\bm E} \Or{\pm}_i \Ol{\pm}_j \ket{\bm E} 
    - 
    \Im(\omega)^2 
    \bra{\bm E} \Or{\mp}_i \prod_{\mathclap{i<k<j}} Z^{A_2}_k \Ol{\mp}_j \ket{\bm E} 
    \nonumber
    \\
    &\qquad \mp 
    \Re(\omega)\Im(\omega) 
    \left(
    \bra{\bm E} \Or{\pm}_i \Ol{\mp}_j \ket{\bm E}_{\sigma^z_{j-}} 
    - 
    \bra{\bm E} \Or{\mp}_i \Ol{\pm}_j \ket{\bm E}_{\sigma^z_{i+}}
    \right) \, .
    \label{SM_eq:relation_2}
\end{align}
The two terms on the second line involve states with a single $\sigma^z$ bond defect inserted in the $\ket{\bm E}$ MPS. We can evaluate these terms explicitly for the $h=0$ wavefunctions, where they factorize exactly (whereas for $h\neq 0$ they factorize only in the limit where $\vert i - j\vert \gg 1$). 
In particular, 
\begin{equation}
    \begin{tikzpicture}[baseline={(0,1)}]
        \draw (-1.5, 0.0)--( 2.5, 0.0);
        \draw (-1.5, 2.0)--( 2.5, 2.0);
        \draw (-1.0, 0.0)--(-1.0, 2.0);
        \draw ( 2.0, 0.0)--( 2.0, 2.0);
        \node[mpo](t) at (-1,1) {$\Or{\pm}$};
        \node[mpo](t) at (-1,0) {\small $\ket{E}$};
        \node[fill=white] at (0,0) {$\dots$};
        \node[fill=white] at (0,2) {$\dots$};
        \draw[fill=white, draw=black] (1.2,0) circle (.3cm) node {\small $\sigma^z$};        
        \node[mpo](t) at (2,1) {$\Ol{\mp}$};
        \node[mpo](t) at (2,0) {\small $\ket{E}$};
        \node[mpo](t) at (-1,2) {\small $\bra{E}$};
        \node[mpo](t) at (2,2) {\small $\bra{E}$};
        \filldraw[fill=white] (+0.65,0) circle (.7mm);
        \filldraw[fill=white] (-0.50,0) circle (.7mm);
        \filldraw[fill=white] (+1.00,2) circle (.7mm);
        \filldraw[fill=white] (-0.50,2) circle (.7mm);
    \end{tikzpicture} 
    \quad
    =
    \quad
    \begin{tikzpicture}[baseline={(0,1)}]
        \draw[rounded corners] (-1,0)--(-.25,0)--(-.25,2)--(-1.75,2)--(-1.75,0)--(-1,0);
        \draw (-1.0, 0.0)--(-1.0, 2);
        \node[mpo](t) at (-1,1) {$\Or{\pm}$};
        \node[mpo](t) at (-1,0) {\small $\ket{E}$};
        \node[mpo](t) at (-1,2) {\small $\bra{E}$};
        \node at (.15,1) {$\times$};
        \draw[rounded corners] (2,0)--(2.75,0)--(2.75,2)--(0.5,2)--(0.5,0)--(2,0);
        \draw[fill=white, draw=black] (1.15,0) circle (.3cm) node {\small $\sigma^z$};        
        \draw (2.0, 0.0)--(2.0, 2);
        \node[mpo](t) at (2,0) {\small $\ket{E}$};
        \node[mpo](t) at (2,1) {$\Ol{\mp}$};
        \node[mpo](t) at (2,2) {\small $\bra{E}$};
        \filldraw[fill=white] (-1.75,1) circle (.7mm);
        \filldraw[fill=white] (-0.25,1) circle (.7mm);
        \filldraw[fill=white] ( 0.50,1) circle (.7mm);
        \filldraw[fill=white] ( 2.75,1) circle (.7mm);
    \end{tikzpicture} 
    =
     i\frac{\Tr[E^{r^{-1}}] \pm \Tr[E^{r}]}{d_E} \times \frac{\Tr[\sigma^z E^r] \mp \Tr[\sigma^z E^{r^{-1}}]}{d_E}
\end{equation}
and
\begin{equation}
    \begin{tikzpicture}[baseline={(0,1)}]
        \draw (-1.5, 0.0)--( 2.5, 0.0);
        \draw (-1.0, 0.0)--(-1.0, 2.0);
        \draw ( 2.0, 0.0)--( 2.0, 2.0);
        \draw (-1.5, 2.0)--( 2.5, 2.0);
        \draw[fill=white, draw=black] (-.25,0) circle (.3cm) node {\small $\sigma^z$};        
        \node[mpo](t) at (-1,1) {$\Or{\mp}$};
        \node[mpo](t) at (-1,0) {\small $\ket{E}$};
        \node[mpo](t) at (2,1) {$\Ol{\pm}$};
        \node[mpo](t) at (2,0) {\small $\ket{E}$};
        \node[mpo](t) at (-1,2) {\small $\bra{E}$};
        \node[mpo](t) at (2,2) {\small $\bra{E}$};
        \node[fill=white] at (.8,0) {$\dots$};
        \node[fill=white] at (.8,2) {$\dots$};
        \filldraw[fill=white] (+0.25,0) circle (.7mm);
        \filldraw[fill=white] (+1.45,0) circle (.7mm);
        \filldraw[fill=white] (+1.45,2) circle (.7mm);
        \filldraw[fill=white] (-0.10,2) circle (.7mm);
    \end{tikzpicture} 
    \quad
    =
    \quad
    \begin{tikzpicture}[baseline={(0,1)}]
        \draw[rounded corners] (0,0)--(.5,0)--(.5,2)--(-1.75,2)--(-1.75,0)--(0,0);
        \draw (-1.0, 0.0)--(-1.0, 2);
        \draw[fill=white, draw=black] (-.1,0) circle (.3cm) node {\small $\sigma^z$};        
        \node[mpo](t) at (-1,1) {$\Or{\mp}$};
        \node[mpo](t) at (-1,0) {\small $\ket{E}$};
        \node[mpo](t) at (-1,2) {\small $\bra{E}$};
        \node at (.875,1) {$\times$};
        \draw[rounded corners] (2,0)--(2.75,0)--(2.75,2)--(1.25,2)--(1.25,0)--(2,0);
        \draw (2.0, 0.0)--(2.0, 2);
        \node[mpo](t) at (2,0) {\small $\ket{E}$};
        \node[mpo](t) at (2,1) {$\Ol{\pm}$};
        \node[mpo](t) at (2,2) {\small $\bra{E}$};
        \filldraw[fill=white] (-1.75,1) circle (.7mm);
        \filldraw[fill=white] ( 0.50,1) circle (.7mm);
        \filldraw[fill=white] ( 1.25,1) circle (.7mm);
        \filldraw[fill=white] ( 2.75,1) circle (.7mm);
    \end{tikzpicture} 
    =
    i\frac{\Tr[E^{r^{-1}}\sigma^z] \mp \Tr[E^{r}\sigma^z]}{d_E} \times \frac{\Tr[E^r] \pm \Tr[E^{r^{-1}}]}{d_E} \, .
\end{equation}
Using Eq. \eqref{SM_eq:Er_Es},
\begin{equation}
    [\sigma^z, E^r] = 0, 
    \quad 
    \Tr[E^r] = \Tr[E^{r^{-1}}] = 2\Re(\omega) ,
    \quad 
    \Tr[E^r \sigma^z] = 2i\,\Im(\omega) ,
    \quad 
    \Tr[E^{r^{-1}} \sigma^z] = -2i\,\Im(\omega),
    \quad
    d_E = 2,
\end{equation}
and substituting into \cref{SM_eq:relation_2}, we obtain for $h=0$
\begin{align}
    \bra{\bm A_1} \Or{+}_i \Ol{+}_j \ket{\bm A_1} 
    &= 
    \Re(\omega)^2 
    \bra{\bm E} \Or{+}_i \Ol{+}_j \ket{\bm E} 
    - 
    \Im(\omega)^2 
    \bra{\bm E} \Or{-}_i \prod_{\mathclap{i<k<j}} Z^{A_2}_k \Ol{-}_j \ket{\bm E} 
     + 8 \Re(\omega)^2 \Im(\omega)^2 \, ,
    \label{SM_eq:good_relation}
    \\
    \bra{\bm A_1} \Or{-}_i \Ol{-}_j \ket{\bm A_1} 
    &= 
    \Re(\omega)^2 
    \bra{\bm E} \Or{-}_i \Ol{-}_j \ket{\bm E} 
    - 
    \Im(\omega)^2 
    \bra{\bm E} \Or{+}_i \prod_{\mathclap{i<k<j}} Z^{A_2}_k \Ol{+}_j \ket{\bm E} \, .
\end{align}
These equations establish explicit relations between two-point functions of the local charged operators and non-local string order parameters in the fixed point limit $h\to 0$.

\end{document}